\pdfoutput=1
\documentclass[10pt, journal, compsoc, twoside, hidelinks]{IEEEtran}
\usepackage[utf8]{inputenc}
\usepackage[T1]{fontenc}
\usepackage[english]{babel}

\usepackage[nocompress]{cite}
\newcommand{\ignore}[1]{}
\usepackage{fancyhdr}
\usepackage[normalem]{ulem}
\usepackage[hyphens]{url}
\usepackage{booktabs}
\usepackage{xfrac}
\usepackage{algorithm}
\usepackage{algorithmic}
\usepackage{xspace}
\usepackage{tabularx}
\usepackage{placeins}
\usepackage[para]{threeparttable}
\usepackage{multirow}
\usepackage{multicol}
\usepackage{color}
\usepackage{tikz}
\usepackage[binary-units]{siunitx}
\usepackage{amssymb}
\usepackage{pifont}
\usepackage[acronym]{glossaries}
\usepackage[inline]{enumitem}
\usepackage{colortbl}
\usepackage{amsmath}
\usepackage{hyperref}
\usepackage{listings}
\usepackage{etoolbox}
\usepackage[printwatermark]{xwatermark}
\usepackage{mwe}
\usepackage{mathtools}

\def\reviewpass{7}
\def\arxiv{1}

\definecolor{accent1}{HTML}{204a87}
\definecolor{accent2}{HTML}{ce5c00}
\definecolor{accent3}{HTML}{5c3566}

\newcommand{\ubold}{\fontseries{b}\selectfont}
\robustify\ubold
\newcolumntype{k}[1]{S[
  table-format=#1,
  detect-weight,
  input-symbols={()},
]}

\newcommand{\todo}[2]{%
  \ifnum\reviewpass>#1{%
    \unskip\ignorespaces%
  }\else{%
    {\color{red}{\bf [#2]}}%
  }\fi%
}

\definecolor{revision}{rgb}{0.0, 0.4, 0.8}
\def\revcolor{revision}
\if\arxiv0

\newenvironment{revised}{\color{\revcolor}\nopagebreak[4]}{}
\else

\newenvironment{revised}{}{}
\fi

\newcommand{\secref}[1]{{Section~\ref{#1}}}

\newcommand{\figref}[1]{{Figure~\ref{#1}}}
\newcommand{\tabref}[1]{{Table~\ref{#1}}}

\DeclareSIUnit\op{op}

\DeclareSIUnit\OP{op}
\DeclareSIUnit\OPs{\OP\per\s}
\DeclareSIUnit\OPsW{\OP\per\s\per\watt}
\DeclareSIUnit\GOPs{\giga\OPs}
\DeclareSIUnit\GOPsW{\giga\OPsW}
\DeclareSIUnit\TOPs{\tera\OPs}
\DeclareSIUnit\TOPsW{\tera\OPsW}

\DeclareSIUnit\FLOP{flop}
\DeclareSIUnit\FLOPs{\FLOP\per\s}
\DeclareSIUnit\FLOPsW{\FLOP\per\s\per\watt}
\DeclareSIUnit\GFLOPs{\giga\FLOPs}
\DeclareSIUnit\GFLOPsW{\giga\FLOPsW}
\DeclareSIUnit\TFLOPs{\tera\FLOPs}
\DeclareSIUnit\TFLOPsW{\tera\FLOPsW}

\DeclareSIUnit\cycle{cycle}
\DeclareSIUnit\word{word}

\DeclareSIUnit\GE{GE}
\DeclareSIUnit\kGE{\kilo\GE}
\DeclareSIUnit\MGE{\mega\GE}

\graphicspath{{figures/}}

\newcolumntype{C}{>{\centering\arraybackslash}X}
\newcolumntype{R}{>{\raggedleft\arraybackslash}X}
\newcolumntype{x}[1]{>{\centering\arraybackslash\hspace{0pt}}p{#1}}

\lstdefinelanguage{rvasm}{
  sensitive = true, % case sensitive
  morecomment = [l]{;}, % line comment starts with ';'
}

\newacronym{agu}{AGU}{Address Generation Unit}
\newacronym{cisc}{CISC}{Complex Instruction Set Computer}
\newacronym{csr}{CSR}{Control and Status Register}
\newacronym{dag}{DAG}{Data Address Generator}
\newacronym{dsp}{DSP}{Digital Signal Processing}
\newacronym{fifo}{FIFO}{First-In First-Out}
\newacronym{fma}{FMA}{Fused Multiply-Add}
\newacronym{fpu}{FPU}{Floating Point Unit}
\newacronym{fp}{FP}{Floating Point}
\newacronym{hpc}{HPC}{High Performance Computing}
\newacronym{isa}{ISA}{Instruction Set Architecture}
\newacronym{lsu}{LSU}{Load-Store Unit}
\newacronym{risc}{RISC}{Reduced Instruction Set Computer}
\newacronym{simd}{SIMD}{Single Instruction Multiple Data}
\newacronym{simt}{SIMT}{Single Instruction Multiple Data}
\newacronym{sm}{SM}{Streaming Multiprocessor}
\newacronym{ssa}{SSA}{Single Static Assignment}
\newacronym{ssr}{SSR}{Stream Semantic Register}
\newacronym{tcdm}{TCDM}{Tightly Coupled Data Memory}
\newacronym{tdp}{TDP}{Thermal Design Power}
\newacronym{vliw}{VLIW}{Very Long Instruction Word}
\newacronym{vpu}{VPU}{Vector Processing Unit}
\newacronym{vrf}{VRF}{Vector Register File}

\begin{document}

%%%%%%%%%%%%%%%%%%%%%%%%%%%%%%%%%%
%%   Title, Authors, Abstract   %%
%%%%%%%%%%%%%%%%%%%%%%%%%%%%%%%%%%

% \def\PaperTitle{Mitigating the von Neumann Bottleneck in Single-Issue Cores with Stream Semantic Registers}
% \def\PaperTitle{Stream Semantic Registers: A Lightweight RISC-V ISA Extension to Achieve Full Utilization in Single-Issue Cores on Compute-Intensive Workloads}
\title{Stream Semantic Registers: A Lightweight RISC-V ISA Extension Achieving Full Compute Utilization in Single-Issue Cores}

\author{Fabian~Schuiki,
  Florian~Zaruba,
  Torsten~Hoefler,
  and~Luca~Benini% <-this % stops a space
  \IEEEcompsocitemizethanks{
    \IEEEcompsocthanksitem F. Schuiki, F. Zaruba, and L. Benini are with the Integrated Systems Laboratory (IIS), Swiss Federal Institute of Technology, Zurich, Switzerland. E-mail: \{fschuiki,zarubaf,lbenini\}@iis.ee.ethz.ch
    \IEEEcompsocthanksitem T. Hoefler is with the Scalable Parallel Computing Laboratory (SPCL), Swiss Federal Institute of Technology, Zurich, Switzerland. E-mail: htor@inf.ethz.ch
    \IEEEcompsocthanksitem L. Benini is also with the Department of Electrical, Electronic, and Information Engineering (DEI), University of Bologna, Bologna, Italy.
  }
  % \if\arxiv0
  %   \thanks{Manuscript received November 15, 2019; revised March 30, 2020.}
  % \fi
}

\markboth{
  \if\arxiv0
    IEEE Transactions on Computers,~Vol.~{(vol)}, No.~{(no)}, {(month)}~{(year)}
  \fi
}{%
  Schuiki \MakeLowercase{\textit{et al.}}: Stream Semantic Registers
}

\IEEEtitleabstractindextext{%
\begin{abstract}
Single-issue processor cores are very energy efficient but suffer from the von Neumann bottleneck, in that they must explicitly fetch and issue the loads/storse necessary to feed their ALU/FPU.
Each instruction spent on moving data is a cycle not spent on computation, limiting ALU/FPU utilization to 33\% on reductions.
We propose "Stream Semantic Registers" to boost utilization and increase energy efficiency.
SSR is a lightweight, non-invasive RISC-V ISA extension which implicitly encodes memory accesses as register reads/writes, eliminating a large number of loads/stores.
We implement the proposed extension in the RTL of an existing multi-core cluster and synthesize the design for a modern 22nm technology.
Our extension provides a significant, 2x to 5x, architectural speedup across different kernels at a small 11\% increase in core area.
Sequential code runs 3x faster on a single core, and 3x fewer cores are needed in a cluster to achieve the same performance.
The utilization increase to almost 100\% in leads to a 2x energy efficiency improvement in a multi-core cluster.
The extension reduces instruction fetches by up to 3.5x and instruction cache power consumption by up to 5.6x.
Compilers can automatically map loop nests to SSRs, making the changes transparent to the programmer.
\end{abstract}

}

\maketitle

\IEEEdisplaynontitleabstractindextext

\IEEEpeerreviewmaketitle

%%%%%%%%%%%%%%%%%
%%   Content   %%
%%%%%%%%%%%%%%%%%

% ==============================================================================
\IEEEraisesectionheading{\section{Introduction}\label{sec:intro}}

% Here is a problem.

% Luca: Need a more general introduction paragraph on moore's law slowing down, the power wall and the research focus moving from technology to technology-aware architecture development and architectural innovation with a focus on energy efficiency (i.e. performance at a low power cost).

\IEEEPARstart{T}{he breakdown} of Dennard scaling in modern silicon manufacturing has prompted a paradigm shift in the way we approach computer architecture design.
Cutting-edge processing systems such as today's CPUs and GPUs are hitting the utilization wall \cite{taylor2012dark}.
Research effort is moving away from manufacturing technologies towards technology-aware computer architectures with a focus on energy efficiency.
Performance at low power is a key ingredient in achieving high utilization of available hardware in order to mitigate the effect of limited frequency and overcome dark silicon \cite{pagani2016thermal}.

% \todo3{Dennard scaling; Power wall; Research focus shifting to tech-aware architectures; Focus on energy efficiency; Focus on high utilization of available hardware; Dark silicon; "Utilization wall" \cite{taylor2012dark}}

In-order processor cores built around the load/store paradigm face an efficiency challenge in keeping their functional units busy.
As an illustrative example, assume that we would like to compute the dot product over a long vector of \gls{fp} values.
Consider the following snippet of RISC-V \cite{waterman2017riscv} assembly executed for every pair of values:
%
% \begin{lstlisting}[language=rvasm]
% flw      ft0, 4(a0!)
% flw      ft1, 4(a1!)
% fmadd.s  ft2, ft0, ft1, ft2
% \end{lstlisting}

\vspace{10pt}
\begin{minipage}{0.8\linewidth}\ttfamily
{\bfseries\color{accent1}flw} {\color{accent3}ft0}, {\color{accent2}4}(a0!) \\
{\bfseries\color{accent1}flw} {\color{accent3}ft1}, {\color{accent2}4}(a1!) \\
{\bfseries\color{accent1}fmadd.s} ft2, {\color{accent3}ft0}, {\color{accent3}ft1}, ft2
\end{minipage}
\vspace{10pt}

\noindent The processor loads two values from memory (\texttt{flw}), then multiplies them and accumulates the result (\texttt{fmadd.s}).
We assume that iteration and pointer adjustment are performed via hardware loops and post-load increment, respectively, which are quite affordable microarchitectural enhancements \cite{gautschi2017near}.
In a single-issue core this takes at least three cycles to execute.
Since only every third instruction performs actual computation, the \gls{fpu} is utilized at most 33\,\% of the time.
The fundamental problem is that in a load-store architecture, data transfers from and to memory have to be encoded explicitly as instructions.
If the machine is only able to issue a single instruction per cycle, every load and store introduces at least one idle cycle in the \gls{fpu}.
The lower a kernel's operational intensity, the more pronounced this problem becomes.
% The problem becomes more pronounced for kernels with lower operational intensities, where more data is needed per operation.

% It's an interesting problem.

One would like to keep a processor's functional units as busy as possible for various reasons. Consider the following two scenarios:

\begin{enumerate}
  \item In the near-threshold regime or at the operating temperatures found in \gls{hpc} or the data center, leakage currents are a significant contributor to power consumption.
  We observe for example a 6.5$\times$ leakage increase from 5\% at \SI{25}{\celsius} to 25\% at \SI{85}{\celsius} in the 22\,nm technology used throughout this paper.
  An idle unit still dissipates leakage power, unless it is power-gated at fine granularity, which is hard to do.
  This leakage power adds to the energy consumed per computation, decreasing overall energy-efficiency.
  \item In an area-constrained setting such as an embedded application, idle cycles can be costly performance-wise.
  A core that keeps its \gls{fpu} busy 50\,\% of the time achieves half of the performance within a given area budget, and conversely requires twice the area for a given performance target, than a core which keeps its \gls{fpu} busy 100\,\% of the time.
\end{enumerate}

% It's an unsolved problem.

Achieving high FPU/ALU utilization is the ultimate goal of a micro-architecture design.
Hence, solutions to this problem exist.
For example a core that can issue multiple instructions per cycle does not have this fundamental limitation, as it can keep the \gls{lsu} and the \gls{fpu} busy at the same time.
In our previous example, such a core would need to be able to issue two loads and one \gls{fp} operation, hence fetching and decoding three instructions, per cycle.
Accommodating such an increase in fetch bandwidth does not come for free \cite{sohi1990instruction}.
In fact, moving to superscalar out-of-order execution requires the instruction fetch interface to at least triple in width, followed by three replicated instruction decoders.
In addition, the parallel execution of \gls{lsu} and \gls{fpu} in non-VLIW cores requires at least some form of dependency tracking and reordering, and is likely to entail register renaming as well.

As a result, achieving high execution unit utilization with out-of-order superscalar execution is not energy efficient \cite{azizi2010energy}.
This is true even for moderately complex multiple-issue cores.
\Gls{cisc} machines can allow for memory accesses to be encoded in a compute instruction directly, thus potentially reducing the bandwidth required to issue three instructions in parallel.
Decoding \gls{cisc} instructions is highly non-trivial however and compiler and micro-architecture become more complex.
As a result, even \gls{cisc} \glspl{isa} such as x86 internally unpack these complex instructions into multiple RISC-like micro-operations.
Other approaches such as vector processors and \gls{vliw} require moving to a much more complex data-parallel micro-architecture and/or incur additional instruction bandwidth, with a major impact on the \gls{isa} or the compiler.
% Decoding \gls{cisc} instructions is highly non-trivial however and likely to come at a comparable hardware cost.

% Here is my idea.

In this paper we propose \glspl{ssr}, a simple and lightweight extension to single-issue load-store architectures to remove its utilization bound without major impacts on the complexity of the micro-architecture and its silicon implementation.
The key idea is to allow loads and stores to be encoded in any instruction for instruction sequences with regular data accesses, instead of explicit load/store instructions.
We do this by giving a few registers stream semantics: reading from or writing to the register issues a read or write into the memory system, respectively.
An \gls{agu} placed outside the processor core allows these memory accesses to follow a programmable affine address pattern which is very common for many compute-intensive workloads such as \gls{dsp}, stencils, and machine learning.
A key advantage of following such a predictable address pattern is the fact that data can be loaded pro-actively, implementing a form of prefetching.
This allows our introductory example to be rewritten as
%
% \begin{lstlisting}[language=rvasm]
% fmadd.s  ft2, ft0, ft1, ft2
% \end{lstlisting}

\vspace{10pt}
\begin{minipage}{0.8\linewidth}\ttfamily
{\bfseries\color{accent1}fmadd.s} ft2, {\color{accent3}ft0}, {\color{accent3}ft1}, ft2
\end{minipage}
\vspace{10pt}

\noindent where \texttt{ft0} and \texttt{ft1} now have stream semantics.
Intuitively the expected speedup is 3$\times$.
We will show that in general the achieved speedup is related to the operational intensity of a kernel, the size of the register file, and the size of the first level memory, and may in practice vary between 2$\times$ and 5$\times$, with speedups on realistic kernels as high as 3.7$\times$.
To sum up, the key contributions of this paper are:

\begin{enumerate}
  % - SSR registers
  % - Data mover and embedding in scratch pad memories
  % - PPA impact analysis and comparison in 22FDX
  % - Compiler integration strategy
  \item A register file extension for \gls{risc} architectures that allows for implicit encoding of data transfers within any instruction. We present the necessary architectural changes in an existing multi-core platform (\secref{sec:arch}).
  \item Competitive experimental results for the proposed architecture applied to a low power multi-core system \cite{gautschi2017near}. We provide a performance, power, and area analysis for an implementation in a 22\,nm technology and compare against other systems (\secref{sec:results}).
  \item A discussion and preliminary results on how to extend compilers to directly emit code that leverages the proposed architectural changes (\secref{sec:results}).
\end{enumerate}

The remainder of this paper is organized as follows: \secref{sec:arch} describes the proposed \gls{ssr} extension, \secref{sec:results} presents an evaluation, experimental results, and comparison to other systems. The remaining sections describe related and future work, and offer concluding remarks.

% ==============================================================================
\section{Architecture}
\label{sec:arch}

We start with a description of the \gls{ssr} extension in detail and show the architectural changes necessary. As an implementation we extend a RI5CY (\emph{``riscy''}) core \cite{gautschi2017near} and PULP cluster \cite{rossi2017energy} with two \gls{ssr} data movers. The RI5CY core has a single-issue in-order pipeline, but it features a rich set of micro-architectural enhancements that make it a challenging baseline when assessing efficiency in executing \gls{dsp} workloads. This includes the support for hardware loops and post-increment load/store operations, which removes the need for branches and many address calculations in the innermost hot loops of a computation: hence RI5CY actually achieves the single-issue performance bound for many kernels. We first provide an overview of the \gls{ssr} architecture and then outline the necessary changes inside the core (\secref{sec:arch_core}), at the core interfaces (\secref{sec:arch_dm}), and in the memory system (\secref{sec:arch_mem}).

% ------------------------------------------------------------------------------
\subsection{Overview}

% Explain division into reg-to-stream part inside the core, and stream-to-mem part outside the core.
% High level overview.

The key idea of \gls{ssr} is to intercept accesses to certain registers at the register file and route those accesses out of the core and into the memory system.
A separate address generator assigns an address to each such access.
Since the address generator is configured up front, this can be seen as proactively prefetching the next accesses, rather than reactively hiding access latency.
In our implementation we assign stream semantics to the \texttt{t0} and \texttt{t1} integer and the \texttt{ft0} and \texttt{ft1} floating point registers.
Reads from and writes to these four registers will be diverted out of the processor.
These are the first two caller-saved integer and floating point registers in RISC-V.

The architectural changes can be subdivided into two main parts: the mapping from register accesses to transactions on a stream interface, and the mapping from those transactions to memory accesses. In our case we perform the mapping to the stream interface inside the RI5CY core and the mapping to memory accesses using a data mover outside the core. This minimizes the changes to RI5CY and exposes a clean interface.
%\todo3{Maybe add figure showing a schematic.}

% ------------------------------------------------------------------------------
\subsection{Core}
\label{sec:arch_core}

% How did we change the core on a high level?
% - Show a pipeline diagram of the core with the added connection highlighted.

The following modifications to the processor core are necessary to map register accesses onto a stream interface:
\begin{itemize}
	\item The register file must be extended to intercept and re-route accesses to a subset of registers.
	\item The stream interfaces, one per register file port, need to be exposed in the core's port list.
	\item A \gls{csr} is needed to enable or disable stream semantics.
	\item Additional stall conditions and back pressure paths introduced by the stream interface must be considered in the core controller.
\end{itemize}
In the following paragraphs we provide a detailed description of each of these modifications.
%\todo3{Maybe add figure showing pipeline diagram with added signals and back-pressure paths.}

\subsubsection{Register File}
\label{sec:arch_regfile}

% What changed in the regfile?
% - 6 bit addr; 5 from RV32, but we fuse the float regs in here
% - high-level data flow view
% - onehot address bitmasked and checked (available anyway due to addr dec)
% - detailed schematic of read port
% - detailed schematic of write port
% - stall control logic

\begin{figure}
	\centering
	\includegraphics[width=\linewidth]{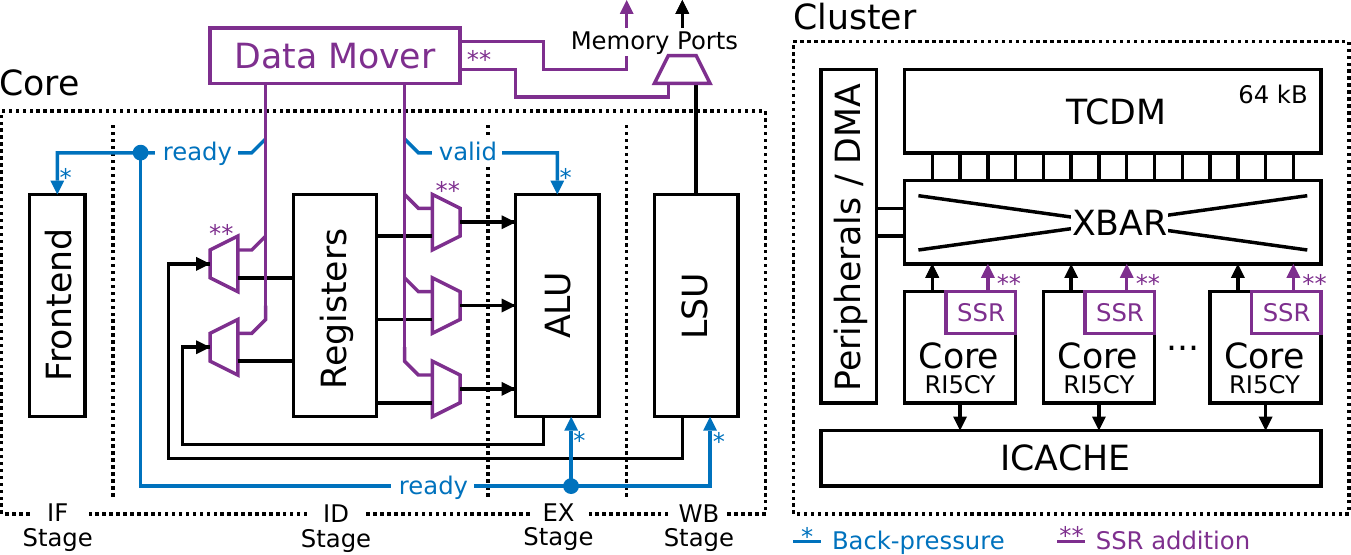}
	\caption{
		On the left: High-level data flow of the \gls{ssr} extension in a single core.
		Read and write accesses to certain registers are filtered and diverted out of the processor core.
		The data mover then assigns memory addresses to the accesses and forwards them into the memory system.
		On the right: cluster of multiple cores, attached to \gls{tcdm} and peripherals via logarithmic interconnect (XBAR).
		Pipeline additions to control back-pressure marked in blue ($\ast$); SSR data path additions marked in purple ($\ast\ast$).
	}
	\label{fig:arch_regfile_dataflow}
\end{figure}

\begin{figure}
	\centering
	\includegraphics[width=\linewidth]{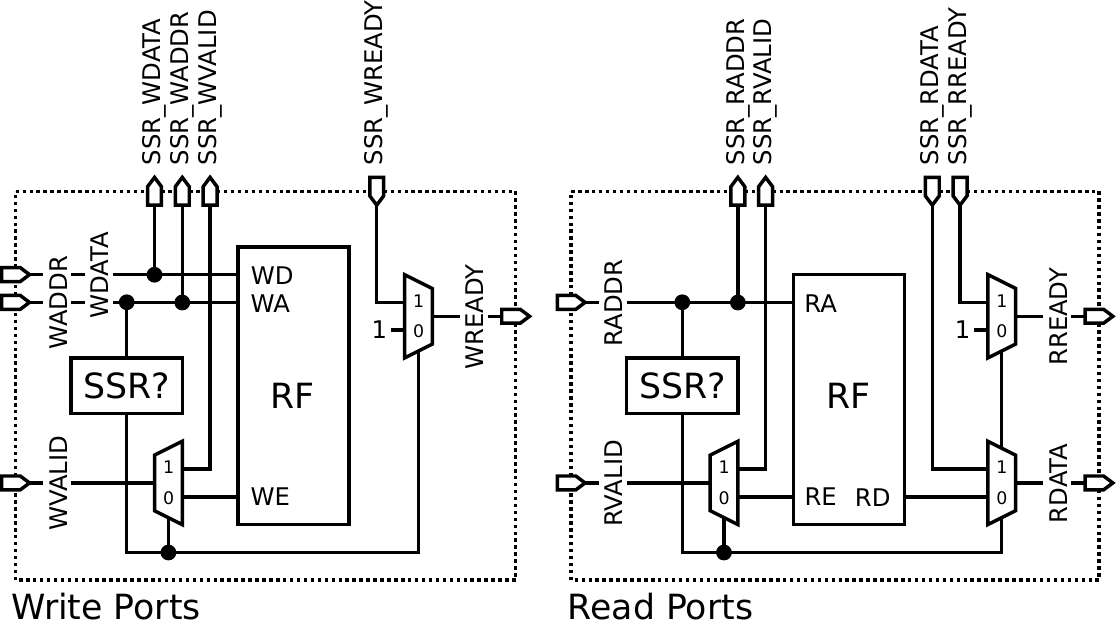}
	\caption{
		Additional circuitry required per register file write port (left) and read port (right) to accomodate stream semantics.
		The additions are implemented as a wrapper around the core's existing register file (RF).
		The \gls{ssr} access check ``SSR?'' is equivalent to evaluating $A \in \{t0,t1,ft0,ft1\} \And E$, where $A$ is the register address and $E$ the \gls{ssr} enable \gls{csr} bit.
		See \secref{sec:arch_regfile} for details.
	}
	\label{fig:arch_regfile}
\end{figure}

The fundamental architectural change of the \gls{ssr} extension is in the processor's register file.
A generic vanilla register file consists of read ports which supply the operands for subsequent pipeline stages, and write ports which store back the result of an instruction.
In our case the original register file has three read and two write ports.
For \gls{ssr} we would like to intercept accesses to a certain set of registers and, instead of accessing the register file, perform a read or write transaction on the external stream interface.
\figref{fig:arch_regfile_dataflow} depicts the high-level data path implied by this architectural change.
Instead of directly routing read and write accesses to the register file, we first determine if an accessed register has stream semantics enabled and if yes, re-route the access onto the corresponding stream interface.
Each port into the register file has a corresponding stream interface.
RISC-V with ``IFD'' extensions allocates 32 integer and 32 float registers, and uses 5\,bit to address them.
The RI5CY core used in our implementation fuses these into a register file with 64 registers and 6\,bit addresses, where the most significant address bit is set depending on whether the requesting instruction is an integer or floating point operation.

\figref{fig:arch_regfile} outlines the additional circuitry needed per write and read port. In both cases we determine if an access has stream semantics (``SSR?'') by checking the following conditions:
\begin{enumerate}
	\item The register address WA or RA must be one of the registers with stream semantics (\texttt{t0}, \texttt{t1}, \texttt{ft0}, \texttt{ft1} in our implementation).
	\item The stream semantics must be enabled in the core's \glspl{csr}.
\end{enumerate}
If both conditions hold the transaction is routed out of the core via a stream interface. These interfaces use a valid/ready handshake which allows for the data mover and memory system to assert back pressure into the core if a request cannot be serviced immediately. The additional hardware is implemented as a wrapper around the core's existing register file.

\subsubsection{Control and Status Registers}
\label{sec:arch_csr}

% What did we change in the CSRs and how is it attached?
% - picked CSR in custom extensions
% - single bit, enables routing of reg accesses to stream ports
% - show CSR layout

The \gls{ssr} extension needs to be opt-in and disabled by default. This allows code that does not benefit from the use of streams to have the full set of registers available. Furthermore it maintains compatibility with existing code and code generated by a compiler that is not aware of the extension. To this end, we have added the \texttt{ssrcfg} \gls{csr} with address \texttt{0x7C0} to the core. It contains a single bit that enables or disables stream semantics in the core. The subset of registers with stream semantics is fixed in hardware and can only be enabled or disabled all at once. Sections of code using \gls{ssr} are expected to set this bit at their beginning and clear it at their end, essentially defining an ``\gls{ssr} region'' in the code. Special care must be taken to handle interrupts and exceptions; see \secref{sec:arch_int}.

\subsubsection{Pipeline Considerations}
\label{sec:arch_pipeline}

Our proposed architectural extension has the following implications on the processor pipeline:
\begin{enumerate}
	\item First and foremost, due to its newly-gained stream nature the register file loses its \emph{idempotency}.
	A non-\gls{ssr} processor may not contain any control signals to precisely enable or disable register accesses during stalls, since reading or writing the same register with identical data has no consequences in an idempotent register file.
	As an example, a multi-cycle instruction may apply bogus data to its destination register during all but the last cycle of execution, and only present valid data in its last cycle.
	As an optimization the processor may choose to enable register writes during all cycles, knowing that the last cycle will eventually override bogus data with the valid result.
	Similarly, an instruction may keep reading registers even during a pipeline stall.
	With \gls{ssr} such optimizations are not possible and additional signals must be added to ensure each instruction accesses its registers exactly once.
	Similarly, a compiler may not delete redundant writes to an \gls{ssr}.
	All of the following considerations are a result of this loss of idempotency.
	\item Instructions that enable or disable stream semantics, CSR writes in the RISC-V case, require pipeline bubbles to be inserted between them and subsequent instructions operating on stream registers.
	This is necessary since the register file access is likely to be located in a different pipeline stage than the CSR write.
	A non-idempotent access to a stream may therefore erroneously occur before a preceding \gls{ssr}-disabling CSR write fully executes.
	Vice versa, a stream access may be skipped due to the preceding \gls{ssr}-enabling write not having taken effect.
	\item Instructions that operate on \glspl{ssr} may not access the register file speculatively.
	In the case of RI5CY, this requires such instructions to be stalled in the decode stage when a branch is currently being resolved.
	\item The register file must be able to exert back-pressure into the pipeline on reads and writes.
	A non-\gls{ssr} processor may not provision for such a scenario.
	In the RI5CY case, additional stall control signals are needed from the register file to the instruction decode stage for reads, and to the write-back stage for writes.
\end{enumerate}

% ------------------------------------------------------------------------------
\subsection{Data Mover}
\label{sec:arch_dm}

\begin{figure}
	\centering
	\includegraphics[width=\linewidth]{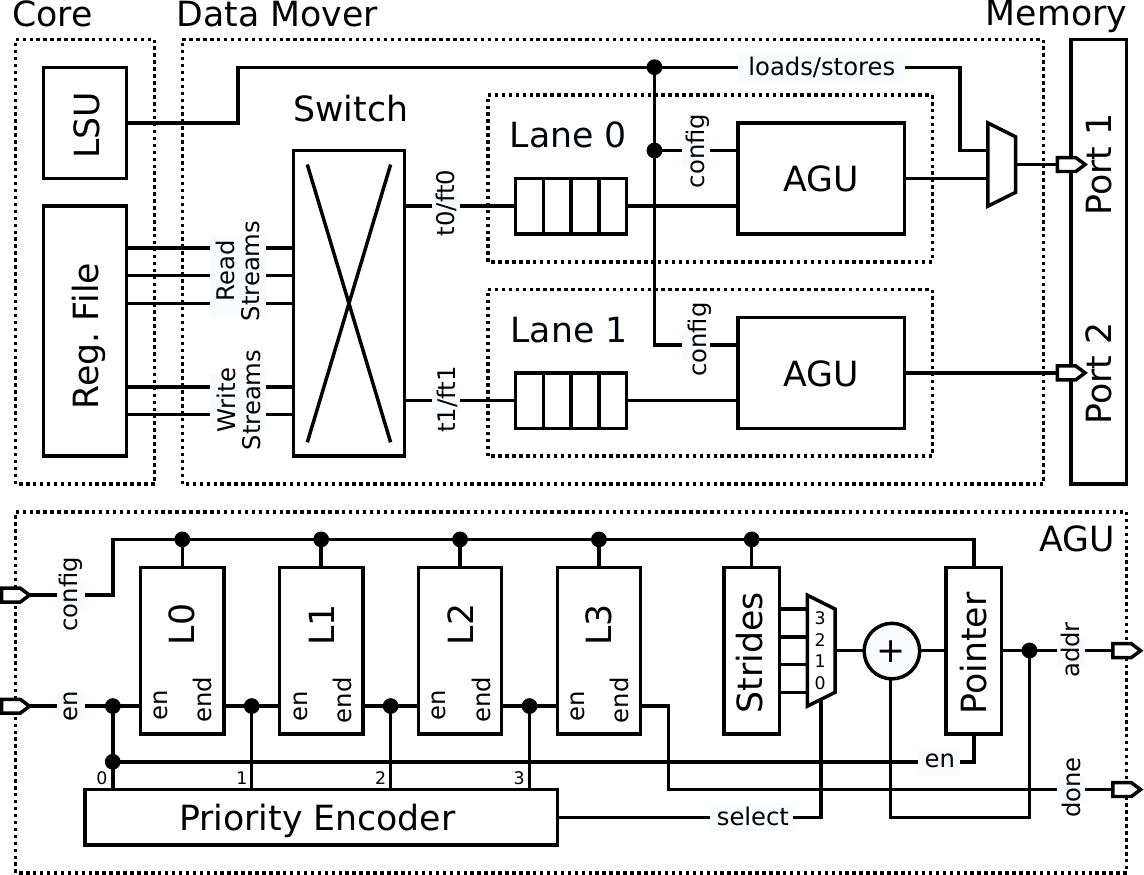}
	\caption{
		Above:
		Architecture of the data mover which translates from a stream of register accesses to the corresponding memory addresses and accesses.
		Below:
		Architecture of an individual \gls{agu} with four nested loops.
		``L0'' to ``L3'' are loop counters.
		Patterns are configured via the ``config'' interface, which modifies the loop counters, strides, and pointer register.
	}
	\label{fig:arch_dm}
\end{figure}

The modifications outlined thus far transport \gls{ssr} register accesses selectively out of the core.
\figref{fig:arch_dm} shows the data mover that is used to map these accesses into the memory system.
It consists of three general parts described in the following.

Each port into the register file is exposed as a separate stream at the boundary of the core, three read and two write streams in our case.
A \emph{switch} uses the register address to map each access on these streams to the targeted data mover lane.
Our example contains two such lanes, one for the \texttt{t0/ft0} and one for the \texttt{t1/ft1} registers.

Each \emph{lane} consists of a \emph{\gls{fifo}} queue to buffer read and write data.
An \emph{address generator} based on the one presented by Schuiki et al. \cite{schuiki2019scalable} and Conti et al. \cite{conti2018xnor} assigns memory addresses to the stream-based accesses performed by the core.
The lane can be put into read mode, in which case the address generator is used to fetch data from memory and store it in the \gls{fifo}.
If put into write mode, the address generator is used to tag each datum written into the \gls{fifo} with an address and send it off into the memory system.
This requires a register to be exclusively used as source or destination operand until the complete address pattern has been exhausted; thus a stream cannot be used to interleave read and write operations.

A significant advantage of this architecture is the fact that the data mover can pro-actively perform memory reads.
In this way, when the processor decides to read from an \gls{ssr}, the datum is already present.
This is in stark contrast to a regular load instruction which merely initiates a memory access.
Subsequent instructions must wait for the accessed data to return, which amounts to at least one cycle of delay.
\gls{ssr}-based operations thus become significantly more memory latency tolerant than regular loads.
This pro-active read introduces a source of incoherence in that a write to a memory location does not update the corresponding value in the read \gls{fifo}.
Read streams started \emph{after} a write will always see the written value.
Thus to avoid data races write operations shall not be performed on a memory range that is currently used in a read stream.
This limitation proves to be inconsequential in all kernels investigated in this paper.

% Note that in general this is not exception and interrupt safe.
% The exact position in each stream must be saved and restored in the interrupt handler, which can be performed in software by accessing the stream configuration registers.
% In our implementation the \glspl{ssr} operates directly on local scratch pad memory without virtual addressing or memory protection, and as such cannot generate load/store exceptions.
% Furthermore since RI5CY is intended to be operated in a many-core setting, we leave interrupt handling to a dedicated core and disable interrupts while \glspl{ssr} are in use.
% We propose to use a coarser-grained approach to interrupt and exception handling as it is found in GPUs.

% ------------------------------------------------------------------------------
\subsection{Interrupts and Exceptions}
\label{sec:arch_int}

\begin{revised}
The \glspl{agu} and enable-state of the \glspl{ssr} add additional implicit state to the processor core that must be adequately saved/restored.
Depending on the features of the processor in question, and the requirements of the target applications, this comes at varying complexity.
In general we identify the following three options of handling exceptions:

\begin{enumerate}
    \item \textbf{No exceptions:}
    The simplest approach is to not support exceptions at all.
    While not applicable to operating-system-capable cores, specialized number-crunching cores, e.g., the ones in a GPU, have very limited exception support \cite{tanasic2017efficient}.
    Since control flow cannot be diverted in this case, no special treatment is required.

    \item \textbf{Deferred exceptions:}
    A similar approach is to defer exception handling to \gls{ssr}-disabled regions.
    While this may incur significant latency in the presence of long-running streams, it allows number-crunching cores to support a minimal amount of external interrupts.
    Applicability is limited to a similar scenario as the above approach, since the utility of deferred exceptions is very limited in practice \cite[p.~261]{ibm2007cell}.
    This is mainly due to difficult debugging and operating systems requiring finer-grained exceptions.
    It is also necessary that the core cannot generate memory exceptions, since streaming over a sufficient number of pages could otherwise livelock the core.

    \item \textbf{Precise exceptions:}
    It is also possible to fully support precise exceptions with \glspl{ssr}.
    This is required for cores with virtual memory and/or memory protection, or cores that must give timing guarantees on exception handling.
    The \glspl{ssr} expose their precise architectural state as registers, which become part of the saved/restored thread context.
    Streams can thus be interrupted and resumed at will.
    Such a scheme is easily implemented in an operating system, that already features a platform-specific exception handler.
    As an upper bound, precise exceptions double the area and power of the \glspl{agu}, since now in addition to the \gls{agu} running ahead as data is requested, a separate \gls{agu} is required which advances in synchrony with the instruction stream, and thus represents the ``architectural state'' of the stream.
    Upon an exception, this \gls{agu} reflects the state before the exception occurred, and is used to re-start execution once the exception handler returns.
\end{enumerate}

\begin{table}
\begin{threeparttable}
\caption{
	Levels of exception support required for different processor core features and applications.
	See \secref{sec:arch_int} for details.
}
\label{tbl:arch_int}
\begin{tabularx}{\linewidth}{@{}Xlll@{}}
\toprule
                           & \textbf{External}   & \textbf{Virtual} & \\
\textbf{Exception Support} & \textbf{Interrupts} & \textbf{Memory}  & \textbf{Application}   \\
\midrule
\textbf{1)} No exceptions       & no  & no  & GPU \\
\textbf{2)} Deferred exceptions & yes & no  & IoT \\
\textbf{3)} Precise exceptions  & yes & yes & CPU \\
\bottomrule
\end{tabularx}
\end{threeparttable}
\end{table}

Consider \tabref{tbl:arch_int} for an overview of processor features, applications, and required exception support.
Our implementation is of the ``deferred exceptions'' kind: the RI5CY core supports external interrupts, but these remain disabled during \gls{ssr} operation.
Since we operate directly on local scratch pad memory without virtual addressing or memory protection, the only source of memory exceptions are accesses to unmapped addresses, after which the program is not expected to resume.
Rather, the exception handler aborts the \gls{ssr} streams and terminates the program.
However, note that this does not preclude the use of virtual memory in the remainder of the system:
data is moved explicitly via a DMA which can support efficient page fault handling \cite{kurth2018scalable}.
\end{revised}

% ------------------------------------------------------------------------------
\subsection{Memory System}
\label{sec:arch_mem}

% How does this change the memory system?
% - multiplexing of core ports
% - banking factor?
% - area cost?

It is important to note that the \gls{ssr} extension increases the peak memory bandwidth that a core can request from the memory system.
In our implementation this translates into an increase in the number of ports into the local scratch pad memory, but may correspond to additional ports into the top-level cache in other systems.
The following aspects are relevant for choosing a reasonable number of \glspl{ssr}, streams, data movers, and ports into memory.

\subsubsection{Impact of SSRs and Streams}
\label{sec:arch_mem_ssr}

On an architectural level, the memory traffic that a core may generate via \glspl{ssr} is limited by two factors:
(1) the number of ports into the register file, and
(2) the number of register operands in an instruction.
The former is five in our implementation (three read and two write ports).
The latter is four as defined by RISC-V \cite{waterman2017riscv} (three source and one destination registers).
The smaller of the two is the upper bound on the memory traffic that can be generated, \SI{4}{\word\per\cycle} in our case.

\subsubsection{Impact of Data Movers}
\label{sec:arch_mem_dm}

The number of data movers determines the number of independent memory address patterns a core can keep track of.
Since the data movers are tied to individual registers, there need to be at least the same number of registers with stream semantics as there are data movers.
Multiple \glspl{ssr} may address the same data mover, for example to use the data mover both in integer and \gls{fp} instructions.
This is the case in our implementation where the \texttt{t0} and \texttt{ft0} registers are bound to the same data mover lane.
As such the number of data movers puts an additional upper bound on the memory traffic a core may generate, \SI{2}{\word\per\cycle} in our implementation.

In this paper we only consider a simple data mover that generates a single memory access per stream transaction.
This is not a requirement.
More elaborate data movers may issue more memory traffic.
Consider indirect addressing for example: in such a setting a data mover would load an address from memory, then use that address to perform the load or store corresponding to a stream transaction, effectively accessing \SI{2}{\word\per\cycle}.
Depending on the cost per memory port and the achieved speedup, such schemes may warrant the use of more than one port per data mover.

Not every data mover necessarily requires a separate port into memory, however.
A system might provide, for example, eight data movers to independently track eight separate sequences of memory addresses, and instructions would pick a sequence via the stream registers they use.
However such a system would still be limited to \SI{4}{\word\per\cycle} of traffic due to the limited number of operand registers (see \secref{sec:arch_mem_ssr}).
In this case it would be beneficial to multiplex the eight data movers onto four memory ports via an arbitration scheme.

\subsubsection{Impact of Memory Ports}
\label{sec:arch_mem_ports}

Due to the nature of single-issue load-store architectures, an instruction may either exercise the \gls{lsu} or the data mover, but never both.
(Technically an \gls{ssr} could be used as the destination register for a load, but we do not consider \texttt{memcpy} a critical application.)
Therefore the cumulative bandwidth generated by the \gls{lsu} and one of the data movers never exceeds \SI{1}{\word\per\cycle}.
Since memory ports are costly we thus suggest to always multiplex the core's \gls{lsu} and one of the data movers into a single port.

% We suggest to consider the operational intensity of a kernel, given by the number of instructions executed per word of data transferred from or to memory, to optimize the overall number of memory ports per core for the expected workload.
Reduced to their fundamental instruction, operations such as multiply-add have an intensity of \SI{0.25}{\op\per\word}, addition or multiplication one of \SI{0.33}{\op\per\word}, and multiply-accumulate of \SI{0.5}{\op\per\word}, meaning for every operation performed they consume and produce four, three, and two data words, respectively.
In order to sustain one instruction per cycle, a core would require four, three, and two ports into the memory system, respectively.
% With the sustained operational intensity of the core decided, we choose the number of memory ports accordingly and allocate enough \glspl{ssr} and data movers to not create additional bottlenecks as described previously.
Our implementation allocates two memory system ports per core and uses two data movers, one of which is multiplexed with the core's LSU data port using a fixed priority arbitration scheme. This allows the core to sustain kernels with an operational intensity of \SI{0.5}{\op\per\word} or higher, which covers the pervasive multiply-accumulate operations found in linear algebra and machine learning.

% ==============================================================================
\section{Programming Model}
\label{sec:progmod}

\begin{figure}
	\centering
	\includegraphics[width=\linewidth]{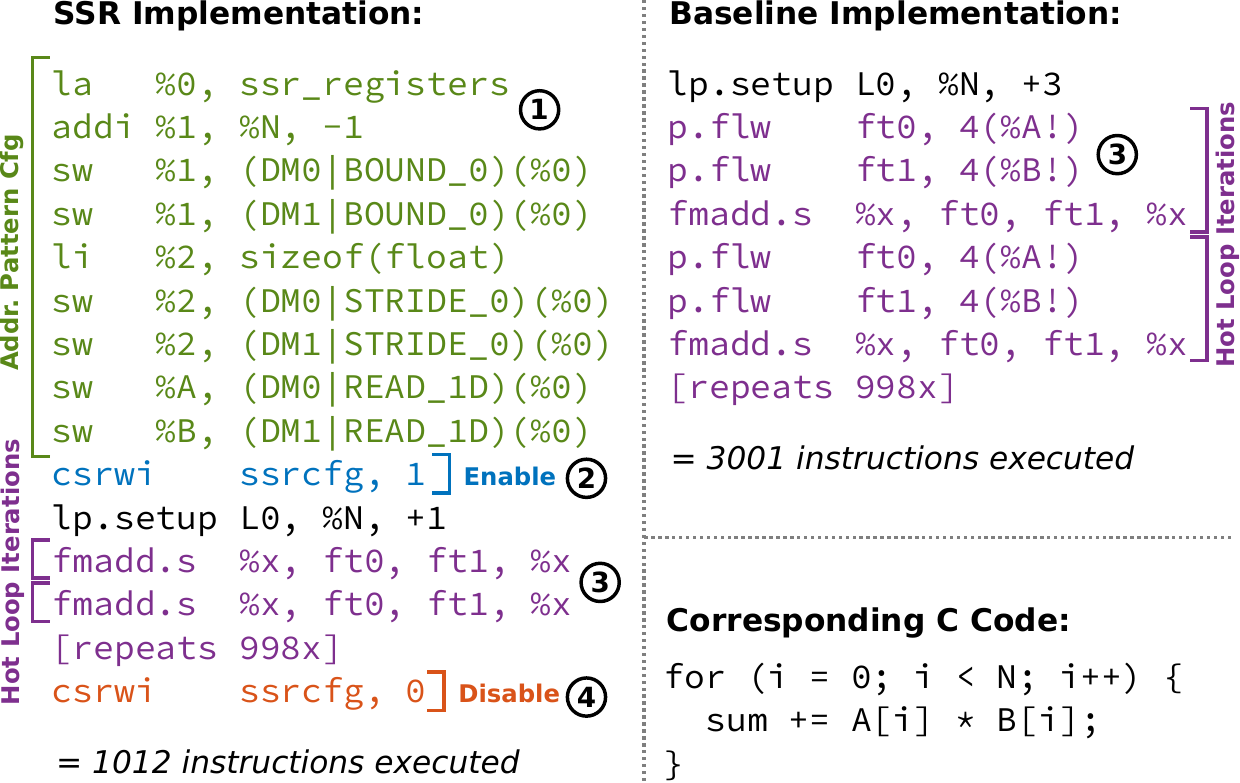}
	\caption{
		Basic usage pattern of \glspl{ssr}.
		Address patterns are configured by writing to the memory mapped address generator registers (1),
		enabling the stream semantics by writing to the \texttt{ssrcfg} \gls{csr} (2),
		performing the actual hot loop (3),
		and disabling the stream semantics again (4).
		Right-hand side shows baseline RISC-V implementation with RI5CY custom extensions.
		Note the two additional post-increment load operations per hot loop iteration absent from the \gls{ssr} case.
		See \secref{sec:progmod}.
	}
	\label{fig:progmod_asm}
\end{figure}

The fundamental \gls{ssr} usage follows the simple sequence outlined in \figref{fig:progmod_asm}: address pattern configuration (1), enabling the stream semantics (2), computation (3), and disabling the stream semantics again (4).
The configuration registers of the data mover are memory mapped and can be accessed by the processor via load and store instructions.
An in-depth explanation of the available registers follows in \secref{sec:progmod_patterns}.
The region of code that makes use of the stream semantics (the ``\gls{ssr} region'') must be surrounded by writes to the \texttt{ssrcfg} CSR to enable \glspl{ssr} upon entry and disable them again upon exit of the region.
The \gls{ssr} region itself can contain any sequence of assembly instructions.

% ------------------------------------------------------------------------------
\subsection{Pattern Configuration}
\label{sec:progmod_patterns}

% \figref{fig:progmod_asm} shows an example of how a one-dimensional summation loop can be mapped to \glspl{ssr}.
% \todo3{Add table with registers.}
Each address generator contains ten configuration registers and supports up to four nested loop dimensions.
We have found four dimensions to cover all problems investigated later in \secref{sec:results_kernels}.
This number is a design parameter and can be changed.
Looping over additional outer dimensions may be performed in software.
The \texttt{status} register contains the address pointer, the number of enabled nested loop dimensions, stream direction (read or write), and a flag indicating whether the end of the pattern has been reached.
A streaming operation is triggered by writing to this register.
The \texttt{repeat} register allows each datum loaded from memory to be emitted into the core multiple times.
This is useful if a value loaded from memory is used as an operand multiple times.
Eight registers control the iteration behavior, two for each loop dimension.
The \texttt{bound0-3} registers contain the number of iterations and the \texttt{stride0-3} registers the address increment for each loop.
% To reduce the number of instructions required to initiate a stream from three to one, the \texttt{pointer} alias registers may be used.
% These are write-only pseudo-registers which write a base address to the \texttt{status} register, implicitly deriving the direction and dimension bits from the register address.
Note that while the \gls{ssr} extension allows for many address stepping and data transfer instructions to be removed from the instruction stream, the program must still issue the exact number of compute instructions (such as \texttt{fmadd}) to fully exhaust the pattern in the address generator.
This means that the fundamental loop nest containing the compute instruction must still be present.
As we will show this is most easily accomplished through the use of hardware loops.

% - [x] Available options
% - [x] 1D to 4D loops
% - [x] Repeat feature
% - [x] Base pointer
% - [x] Aliases to quickly start operation (compare this)

% ------------------------------------------------------------------------------
\subsection{Automated Code Generation in LLVM}
\label{sec:progmod_llvm}

Mapping nested loops from an input language such as C to \glspl{ssr} is straightforward if the loop bounds are constant for the duration of the loop and addresses are a linear function of the indices.
This is the case for many data-oblivious kernels where the control flow does not depend on the data values \cite{goldreich1996software}.
We propose the following recipe for a pass to map loops to \glspl{ssr} in the LLVM compiler framework \cite{lattner2004llvm}.
The pass operates on the Machine IR (MIR) and is executed after instruction selection and before register allocation takes place.
At this stage trivial loop induction variables have been identified and mapped to simple address increments as far as possible.
Our pass is divided into the following phases:

\begin{enumerate}
	\item Identify loops in the MIR data and control flow graph.
	This information is already provided by the LLVM infrastructure.

	\item Visit all load and store instructions within the identified loops.
	Check if the address expression is a simple counter which increments by a constant amount.
	Since the MIR is in \gls{ssa} form this can be done as a simple pattern match:
	We check if the address is determined by a \emph{phi} and \emph{add} node loop in the graph, and whether the input to the \emph{add} is constant.
	This check is done recursively across multiple nested loop levels.
	Loads and stores which pass are marked as candidates for \glspl{ssr} replacement.

	\item Allocate the candidates to the available data movers (two in our case).
	We start with the deepest candidates first, in terms of nesting level, which is a simple heuristic for the number of loop iterations.

	\item Emit instructions to configure the \gls{ssr} before the loop header.

	\item Remove the load/store instruction from the MIR and replace any uses with the corresponding stream register.

	\item Block registers with stream semantics in the register allocation pass.
\end{enumerate}

A limitation of following this recipe without further considerations is that all loops will be mapped to \glspl{ssr} as far as possible.
However, as we will show later, not every loop benefits from \glspl{ssr}.
Especially very short loops may take longer to execute with \glspl{ssr} due to the increased setup overhead.
The decision whether to ``SSR-ify'' a loop should thus be made either at compile time based on the expected number of iterations via an execution trace or heuristic.
Or at runtime based on the actual number of iterations, in which case both an \gls{ssr} and non-\gls{ssr} implementation must be provided.

% ==============================================================================
\section{Performance Analysis}
\label{sec:results}

In this section we evaluate the impact and benefits of \glspl{ssr} on the ISA (\secref{sec:results_isa}), a single core (\secref{sec:results_core}), and an entire cluster (\secref{sec:results_cluster}) in terms of performance, and area and energy efficiency.
% \todo3{Results for 6 core cluster use 6 FPUs at the moment. Either (1) switch to shared FPU, (2) estimate area/power of 3 FPU version based on 6 FPU version, (3) keep current results and make sure text mentions 6 FPUs everywhere.}

% ------------------------------------------------------------------------------
\subsection{ISA-level Impact}
\label{sec:results_isa}

\begin{figure*}
	\centering
	\includegraphics[width=\linewidth]{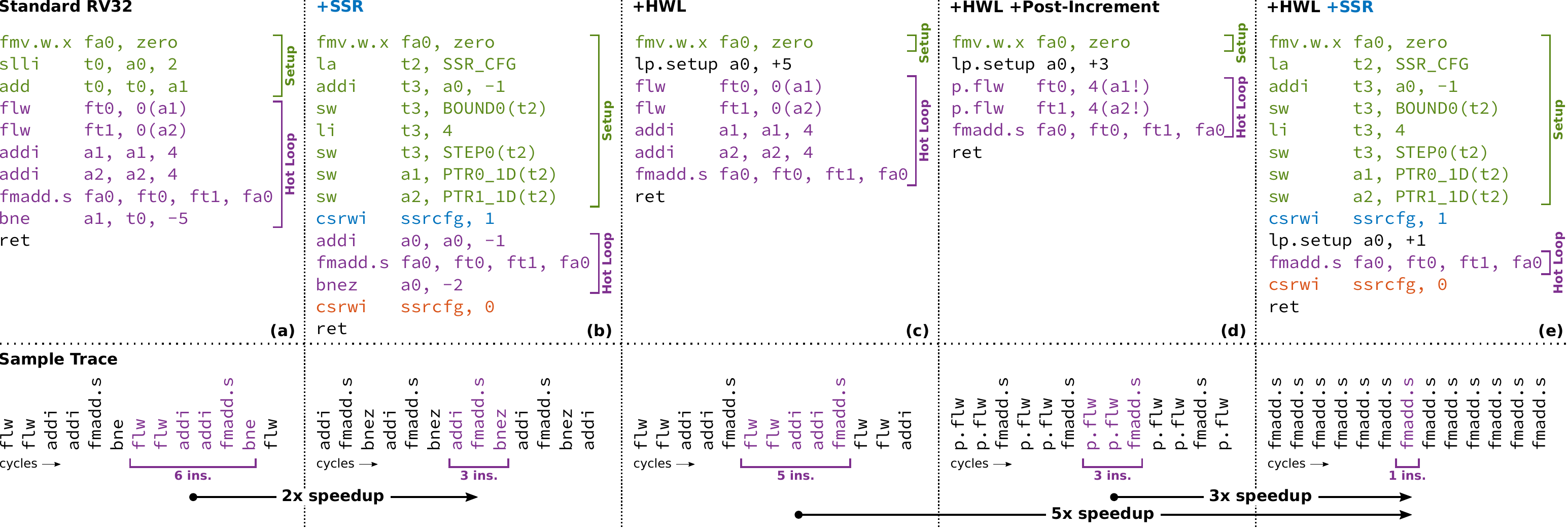}
	\caption{
		Comparison of assembly kernels with different \gls{isa} extensions (above).
		Sample execution trace for each kernel with a single hot loop iteration highlighted (below).
		From left to right: (a) baseline RV32 processor; (b) RV32 with \gls{ssr} extension; (c) RV32 with hardware loops (HWL); (d) RV32 with hardware loops and post-increment loads; (e) RV32 with hardware loops and \gls{ssr} extension.
		See \secref{sec:results_isa} for details.
	}
	\label{fig:results_isa_code}
\end{figure*}

In this section we provide an evaluation of the performance and unit utilization impact of \glspl{ssr} at the \gls{isa} level.
We assume an ideal memory system with a constant access latency of one cycle.
Consider the assembly code for a reduction operation in \figref{fig:results_isa_code}a as an example.
In a standard RISC-V implementation the hot loop consists of six instructions: two loads, two pointer increments, one \gls{fma}, and a branch.
Of these only the \gls{fma} is executed on the \gls{fpu} and performs actual work towards the result, putting the upper bound of \gls{fpu} utilization at 17\%.
The proposed \gls{ssr} extension allows the loads and pointer increments to be implicitly encoded in the use of ft0 and ft1 as input registers, as shown in \figref{fig:results_isa_code}b.
This reduces the number of instructions in the hot loop to three: one counter decrement, one \gls{fma}, and a branch; putting the \gls{fpu} utilization bound at 33\%.
Thus in a standard RISC-V \gls{isa} core the \glspl{ssr} bring an architectural speedup of $2\times$.

\glspl{ssr} interoperate well with hardware loop extensions such as those described by Gautschi et al. \cite{gautschi2017near}.
The use of hardware loops removes the back-branch from the hot loop and alleviates the need to explicitly track a loop counter.
The baseline code shown in \figref{fig:results_isa_code}c now has five instructions in the hot loop, with an \gls{fpu} utilization bound of 20\%.
Using \glspl{ssr} the hot loop reduces to a single instruction: the \gls{fma}, as shown in \figref{fig:results_isa_code}e.
Since loads and pointer adjustments are handled by \gls{ssr}, and the loop iteration is handled by the hardware loop, the only thing left is the actual computation.
This puts the \gls{fpu} utilization to 100\%. The hardware loops introduce an additional \texttt{lp.setup} instruction in the setup code.
Thus in the presence of hardware loops the proposed \gls{ssr} extension brings an architectural speedup of $5\times$ with respect to a vanilla micro-architecture.

The RI5CY core \cite{gautschi2017near} used in our evaluation provides additional instruction set extensions such as load/store post-increment instructions.
These allow for the pointer increments in the baseline code to be elided as shown in \figref{fig:results_isa_code}d, reducing the hot loop to three instructions: two post-increment loads and one \gls{fma}.
Thus the upper bound on \gls{fpu} utilization is at 33\%, and the \gls{ssr} extension still provides a speedup of $3\times$.

\subsubsection{Setup Amortization Analysis}
\label{sec:results_isa_amortization}

\begin{figure}
	\centering
	\includegraphics[width=\linewidth]{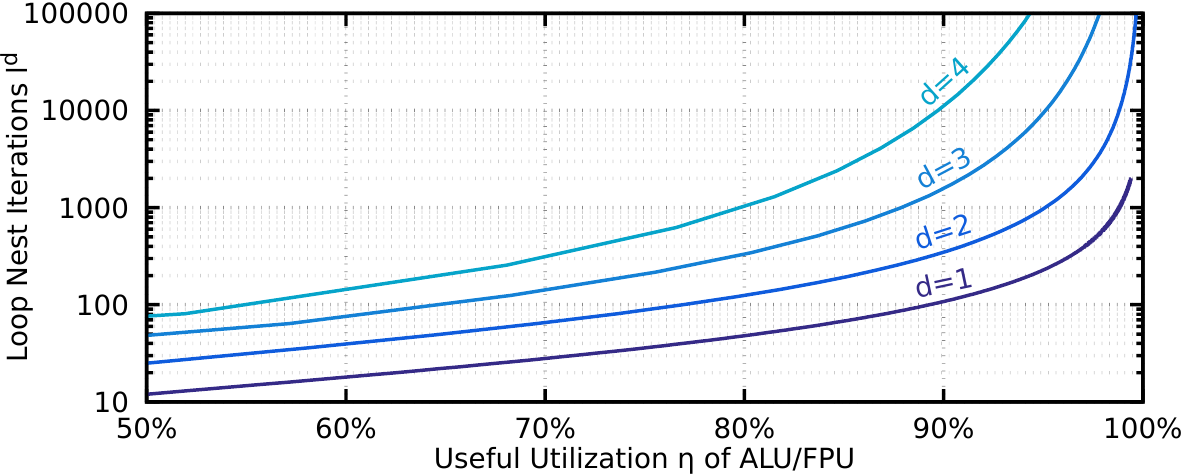}
	\caption{
	Useful ALU/FPU utilization with \glspl{ssr} for a reduction over a $d$-dimensional hypercube with side length $l$, and thus overall number of iterations $l^d$.
	Deeper loop nests contain more overall loop configuration overhead, thus requiring longer-running loops to achieve the same useful utilization as lower-dimensional loops.
	See \secref{sec:results_isa_amortization}.
	}
	\label{fig:results_amortization}
\end{figure}

The \gls{ssr} extension introduces additional setup instructions at the beginning of a loop.
This overhead must be amortized via the speedup gained due to elision of memory transfer instructions in the loop body in order for \gls{ssr} to be beneficial.
For our analysis we assume a loop nest of dimension $d\in\mathbb{N}$ (and corresponding number of hardware loops), $s\in\mathbb{N}$ data movers, and $L\in\mathbb{N}^d$ loop iterations and $I\in\mathbb{N}^d$ instructions per nesting level.
This yields the following model for the total number of executed instructions in the \gls{ssr} and non-\gls{ssr} case:
\begin{align}
N_\text{ssr}  &= \underbracket{4ds+s+2}_\text{(a)} + \sum\limits_{i=1}^d (I_i + 1) \prod\limits_{n=1}^i L_n - \prod\limits_{i=1}^d L_i \label{eqn:results_amortization_ssr} \\
N_\text{base} &= 1 + \sum\limits_{i=1}^d (I_i + \underbracket{1 + s}_\text{(b)}) \prod\limits_{n=1}^i L_n - \prod\limits_{i=1}^d L_i \label{eqn:results_amortization_base}
\end{align}
In the above, (a) describes the one time setup overhead for the \gls{ssr} data movers before the loop nest; and (b) captures the explicit data movement instructions necessary in the non-\gls{ssr} case (for example one load/store per data mover, $s$ in total).
We are now interested in finding the break even point where $N_\text{ssr}$ becomes smaller than $N_\text{base}$ and thus \glspl{ssr} become advantageous. Algebraic transformation yields:
\begin{equation}
N_\text{ssr} \leq N_\text{base} \quad \Rightarrow \quad 4d+2 \leq \sum\limits_{i=1}^d \prod\limits_{n=1}^i L_n
\end{equation}
Interestingly, the number of iterations $I_i$ per nesting level $i$ does not affect the amortization behavior, and neither does the data mover count $s$.
As an instructive example, let us assume that each loop in the nest performs the same number of iterations $l$, and the overall loop nest thus performs $l^d$ iterations.
The \gls{ssr} implementation outperforms the baseline on loop nests with more than 5, 4, 1, or 1 overall iterations $l^d$, for 1D, 2D, 3D, or 4D loop nests, respectively.
For the same scenario, \figref{fig:results_amortization} outlines the achieved useful ALU/FPU utilization $\eta$ with \glspl{ssr} for a reduction over a $d$-dimensional hypercube with side length $l$, and thus overall number of iterations $l^d$.
Each additional loop level in the nest introduces an instruction overhead for loop configuration, therefore requiring exponentially more iterations to achieve the same useful utilization as for lower-dimensional loops.

\subsubsection{Data Dependency Hazards}

\begin{table}
\begin{threeparttable}
\caption{
	Number of instructions $N$, useful ALU/FPU utilization $\eta$, and associated SSR-induced speedup $S$ in the hot loop of a reduction, for integer and floating-point arithmetic, and different numbers of unrolled loop iterations $U$.
}
\label{tbl:results_isa_unrolling}
\begin{tabularx}{\linewidth}{@{}Xlllllll@{}}
\toprule
	\textbf{Kernel} & \textbf{Arith.} & $U$ & \multicolumn{2}{l}{\textbf{no SSR}} & \multicolumn{2}{l}{\textbf{with SSR}} & $S$ \\
	\cmidrule(lr){4-5}
	\cmidrule(lr){6-7}
	&&& $N$ & $\eta$ & $N$ & $\eta$ & \\
\midrule
	Standard RV32    & int32 & 1 & 6 & 17\% & 3 & 33\%  & \textbf{2$\times$} \\
	+ Hardware Loops & int32 & 1 & 5 & 20\% & 1 & 100\% & \textbf{5$\times$} \\
	+ Post-Increment & int32 & 2 & 6 & 33\% & 2 & 100\% & \textbf{3$\times$} \\
\midrule
	Standard RV32    & fp32  & 1 & 6  & 17\% & 3 &  33\% & \textbf{2$\times$} \\
	+ Hardware Loops & fp32  & 3 & 11 & 27\% & 3 & 100\% & \textbf{3.7$\times$} \\
	+ Post-Increment & fp32  & 3 & 9  & 33\% & 3 & 100\% & \textbf{3$\times$} \\
\bottomrule
\end{tabularx}
\end{threeparttable}
\end{table}

Until now we have assumed no data dependency stalls and all instructions to have one cycle latency, which holds for an integer reduction.
Loads and floating point \glspl{fma} have two and three cycles of latency in RI5CY however, which requires some amount of loop unrolling to avoid any stalls. \tabref{tbl:results_isa_unrolling} shows the necessary unrolling and corresponding speedup $S$ provided by \glspl{ssr}.
We define the useful utilization $\eta$ as the fraction of cycles where the computation in the ALU/FPU contributes directly towards the result, as opposed to address or branch calculations.
The integer reduction with post-increment loads requires two-fold loop unrolling in the baseline case to avoid data dependency stalls on the loads.
The \gls{fp} reduction additionally requires three-fold loop unrolling in the SSR case to avoid data dependency stalls on the accumulation register due to the \gls{fma} latency.
If all loop iterations are fully unrolled (no data dependency stalls and all branch and pointer increment instructions fully amortized) the \gls{ssr}-induced speedup reaches $3\times$ in the limit in all cases.

We conclude that our proposed extension provides a significant $2\times$ to $5\times$ speedup, specifically $3\times$ in our implementation in a \gls{dsp}-optimized core.
Furthermore it raises the upper bound for useful compute unit utilization to >95\% for reasonably-sized loop nests.
As we will show in the following sections, these gains are not purely theoretical but translate well into real-life speedups in single- and multi-core settings.

% ------------------------------------------------------------------------------
\subsection{Kernels}
\label{sec:results_kernels}

We have implemented the following kernels on our architecture to evaluate the performance benefits of \glspl{ssr}:
\begin{itemize}
	\item a \emph{reduction} (dot product) over 2048 values;
	\item a \emph{scan} (all prefix sums) over 4096 values;
	\item star-shaped \emph{stencils} as they are found in the discrete Laplace operators (stencil diameter 11) in 1D (1024 points) and 2D (64$\times$64 points);
	\item the dense matrix-vector product (\emph{GEMV}) of a 64$\times$64 matrix and vector of 64 elements;
	\item the dense matrix-matrix product (\emph{GEMM}) of two 32$\times$32 matrices;
	\item the \emph{ReLU} operation found in deep learning ($\text{max}(0,x)$) over 1024 values;
	\item the \emph{fast Fourier transform} over 2048 values; and
	\item a \emph{bitonic sort network} over 1024 values.
\end{itemize}
Problem sizes were chosen to fit into the first-level memory (\gls{tcdm}) to remove effects of DMA transfers to other parts of the memory hierarchy from the analysis.

% ------------------------------------------------------------------------------
\section{Design and Implementation Results}
\subsection{Methodology}
\label{sec:results_methodology}

We based our implementation on RI5CY and the PULP platform \cite{gautschi2017near}, a silicon-proven open-source RISC-V many-core platform written in SystemVerilog.
Our extensions were directly implemented in the core's RTL description.
We use the Synopsys Design Compiler to synthesize individual RI5CY cores and entire PULP clusters with and without the proposed hardware extension.
Synthesis is performed under worst-case conditions at \SI{0.72}{\volt} and \SI{-40}{\celsius} for \textsc{Globalfoundries' 22fdx} technology, a 22\,nm FD-SOI node.
The design is constrained for a \SI{500}{\MHz} worst-case performance target.
To estimate power consumption of the kernels we run simulations of the synthesized netlists in QuestaSim and perform a power estimation on the resulting trace using Synopsys PrimeTime.
Architectural performance improvements in terms of cycles or unit utilization are technology independent.

% ------------------------------------------------------------------------------
\subsection{Single-Core Comparison}
\label{sec:results_core}

\subsubsection{Performance}
\label{sec:results_core_perf}

\begin{figure}
	\centering
	\includegraphics[width=\linewidth]{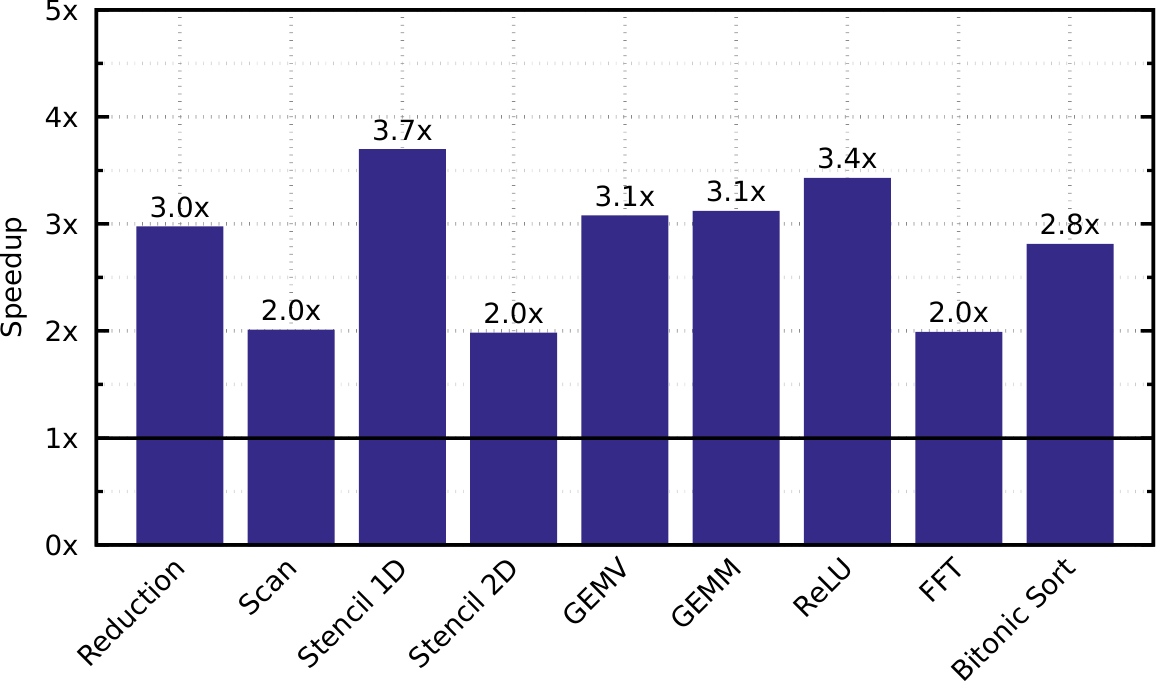}
	\caption{Speedup in a single RI5CY core extended with \glspl{ssr}.}
	\label{fig:results_core_speedup}
\end{figure}

\begin{figure}
	\centering
	\includegraphics[width=\linewidth]{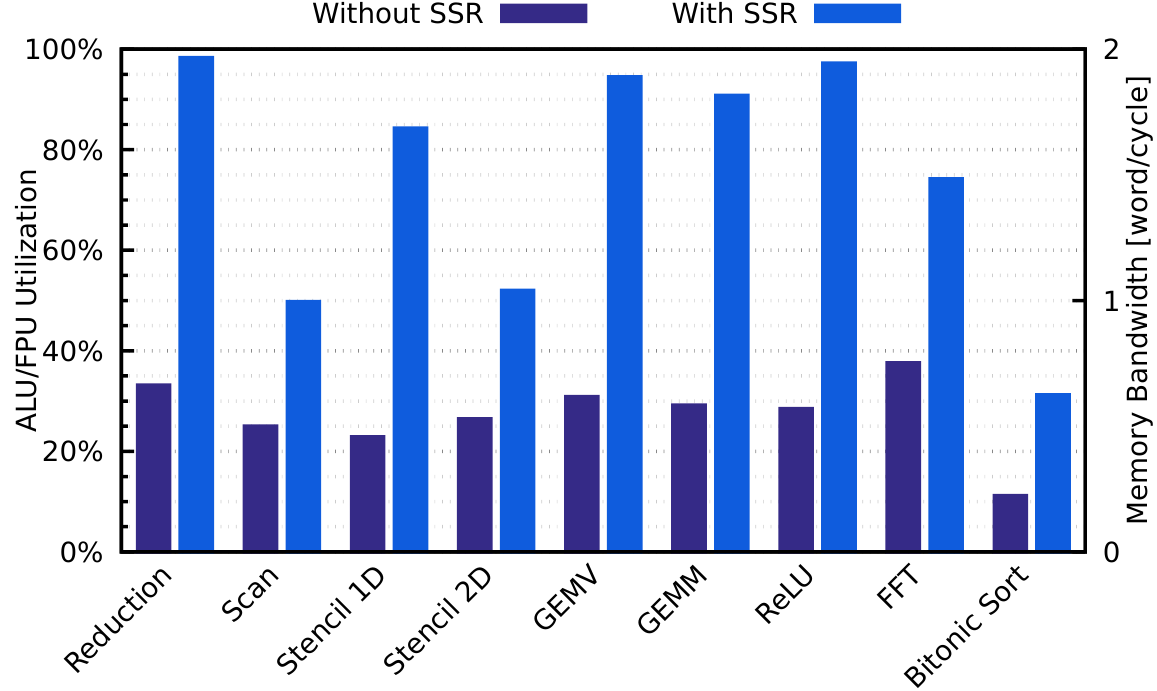}
	\caption{
		Useful ALU/FPU utilization $\eta$ achieved by a single RI5CY core without and with \gls{ssr} extension, including all overheads.
		Note that for the investigated kernels the utilization shown on the left axis is directly proportional to the memory bandwidth shown on the right axis.
	}
	\label{fig:results_core_util}
\end{figure}

\figref{fig:results_core_speedup} shows the speedup achieved by the proposed \gls{ssr} extension in a RI5CY core over a range of data-oblivious kernels.
Again we assume an ideal memory system with a constant access latency of one cycle.
The implementations are fully optimized such that the loop bodies only consists of mandatory non-amortizable instructions.
\glspl{ssr} provide a speedup between 2.0$\times$ and 3.7$\times$ across the kernels, and generally at or above 2$\times$.
% An upper bound on the achievable speedup is given by the ratio of load/store instructions in the hot loop.
% This bound is annotated in \figref{fig:results_core_speedup} \todo3{Actually annotate this}.
% \glspl{ssr} achieve almost perfect speedup across the board.

\figref{fig:results_core_util} shows the useful ALU/FPU utilization $\eta$ of a RI5CY core before and after adding the \gls{ssr} extension when executing the kernels outlined in \secref{sec:results_kernels}.
As outlined in \secref{sec:results_isa}, we consider $\eta$ to be the fraction of instructions executed that contribute to the result.
Without \glspl{ssr}, the utilization is generally around 33\%, heavily bounded by the number of load/store instructions during which the corresponding computational units do not perform any useful work.
With \glspl{ssr}, the hot loop can generally be reduced to the instructions essential for computing the result, such that the utilization during the hot loop reaches close to 100\% in many cases.

Not only does the proposed architecture extension allow for almost perfect utilization of the functional units for many kernels, it also shifts the execution bottleneck from the instruction stream to the data interface.
In the case of, for example, the hot loop of a reduction our implementation without \glspl{ssr} would issue two loads over three cycles (66\% data memory bandwidth utilization with one port), one computation over three cycles (33\% compute utilization), yet three instruction over three cycles (100\% instruction bandwidth utilization).
With \glspl{ssr}, the core issues two loads over one cycle (200\% data memory bandwidth utilization with one port, 100\% with two ports), one computation over one cycle (100\% compute utilization), and one instruction over one cycle (100\% instruction bandwidth utilization).
As indicated in \figref{fig:results_core_util} the achieved memory bandwidth is proportional to the ALU/FPU utilization.
The \glspl{ssr} allow the core to ingest around 3$\times$ more data because a single instruction can consume three registers, which directly corresponds to the speedup and utilization gain achieved.

\subsubsection{Critical Path}
\label{sec:results_core_critical_path}

In our implementation the \gls{ssr} extension adds up to 11 levels of logic to read and write timing paths surrounding the register file.
This includes register address comparison and data multiplexing inside the core, and arbitration and \gls{fifo} access outside of the core.
The critical path of the processor lies within the \gls{fpu} and multiplier in the ``execute'' stage and measures 86 levels of logic.
For comparison the ``instruction decode'' stage where the register file accesses occur covers merely 31 levels of logic.
The addition of 11 levels of logic there has no effect on the critical path.
Additional timing arcs exist from the ``execute'' stage to the register file's write port.
\Glspl{ssr} add 4 levels of logic to these paths, an increase of 4.7\% in terms of logic depth or \SI{85}{\pico\second} (4.3\%) in terms of delay.
We thus conclude that the \gls{ssr} extension increases the critical path by less than 5\%.

\subsubsection{Area}
\label{sec:results_core_area}

\begin{figure}
	\centering
	\includegraphics[width=\linewidth]{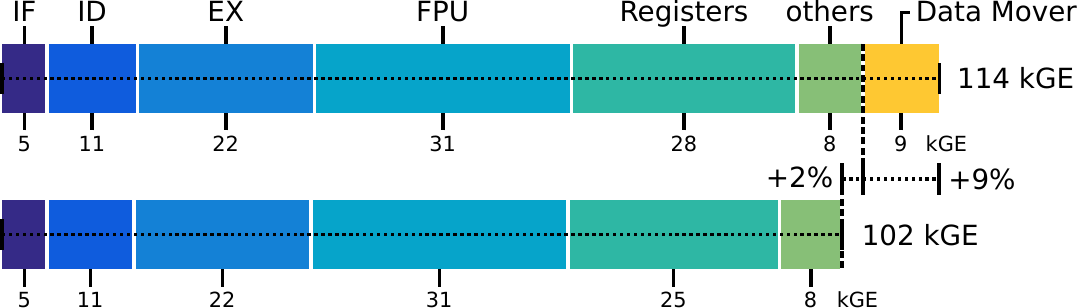}
	\caption{Area breakdown of a RI5CY core with (top) and without (bottom) the \gls{ssr} extension in 22\,nm technology. Shown are the area of the instruction fetch (IF), instruction decode (ID), and execute stages (EX), the FPU, register file, and other components inside the core; as well as the data mover outside of the core. Introduction of \glspl{ssr} amount to a core-internal overhead of 2\,\%, while the addition of a data mover and memory port multiplexer outside of the core increases area by another 9\,\%. In total the addition of \gls{ssr} increases area by 11\,\%.}
	\label{fig:results_core_area_comp}
\end{figure}

\figref{fig:results_core_area_comp} shows the area breakdown of a RI5CY core.
First and foremost, the introduction of \glspl{ssr} increases the area of the RI5CY core by \SI{11.6}{\kGE} or 11\%, of which 2\% are due to core-internal changes around the register file, and 9\% are due to the data mover and memory port multiplexer around the core.
This is a moderate cost considering that a core with \gls{ssr}-equivalent capability would have to issue three instructions per cycle.
For comparison, the SweRV dual-issue core \cite{wd2018swerv} has an area of \SI{237}{\kGE}\footnote{Synthesized by us for the same technology; performance for SSG \SI{0.72}{\volt} \SI{-40/+125}{\celsius} corner. Excludes caches.} at \SI{902}{\MHz}.
This is due to a significant increase in complexity in the instruction fetch and decoding stages, and additional ports into the register file.
Note that this does not yet include the impact on the instruction cache that comes with a tripled instruction bandwidth.
The SweRV core achieves \SI{38.3}{\GOPs\per\milli\meter\squared}, which is 1.7$\times$ less than the \SI{66.1}{\GOPs\per\milli\meter\squared} of RI5CY with \glspl{ssr}.
Note that SweRV is only dual-issue and has no \gls{fp} support; a core with matching triple-issue and \glspl{fpu} would further increase hardware complexity and overhead.

 % extending RI5CY to an \gls{ssr}-equivalent capability of issuing three instructions per cycle would significantly.
% \todo3{Better compare this to SweRV area results by Florian.}
% This is due to at least a threefold replication of the instruction fetch and decoding stages, increase in register file ports, and duplication of the load/store unit.
% As such the real increase in area will be much more pronounced.

\subsubsection{Energy Efficiency}
\label{sec:results_core_eff}

\begin{figure}
	\centering
	\includegraphics[width=\linewidth]{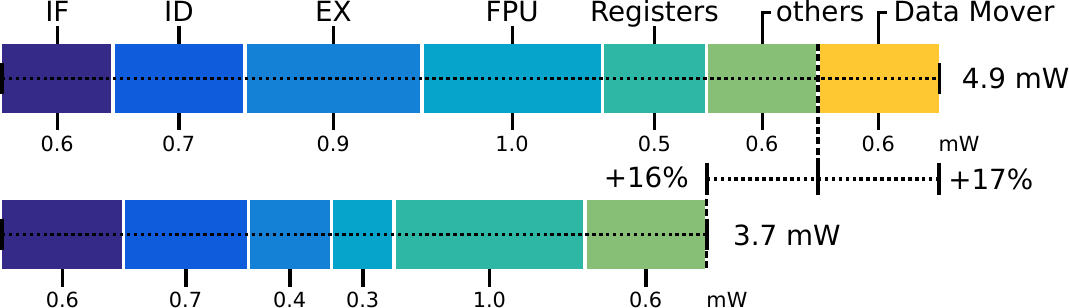}
	\caption{Power breakdown of a RI5CY core with (top) and without (bottom) \glspl{ssr} executing a dot product. \Gls{ssr} accounts for a power consumption increase of 16\% inside and 17\% outside of the core, for a total increase of 33\%. This is due to increased overall activity of the core. Note how \gls{fpu} power increases due to higher utilization, and how register file power decreases due to the \glspl{ssr} directly operating on memory. Power scaled to \SI{1}{\GHz}.}
	\label{fig:results_core_power_comp}
\end{figure}

\figref{fig:results_core_power_comp} shows the power breakdown of a RI5CY core.
The introduction of \glspl{ssr} increases the power consumption of the core by 33\%, of which 16\% are due to core-internal changes and 17\% outside of the core.
Note that the increase in power consumption is more significant than the increase in area, since the addition of \glspl{ssr} significantly increases the computational throughput of the core.
This can be seen in the increased \gls{fpu} power (\SI{0.3}{\milli\watt} to \SI{1.0}{\milli\watt}).
Since most data is now directly streamed from memory, the power consumed by the register file is reduced from \SI{1.0}{\milli\watt} to \SI{0.5}{\milli\watt}, while the new data mover accounts for a corresponding additional \SI{0.6}{\milli\watt}.
Together with the $2\times$ to $3\times$ speedup this translates into an energy efficiency improvement of $1.5\times$ to $2.3\times$ for the various kernels.
Note that this does not yet account for energy saved in the instruction cache due to a reduced instruction bandwidth.
This effect will be explored in the next sections.

% Memory bandwidth?

% This section presents a performance comparison on single cores with ideal instruction and data memory, showing performance and FPU/ALU utilization gains. This allows us to highlight the merits of the architecture on an architectural level, without having to discuss congestion and cache misses already. These will follow in the next section. The following cores will be compared:

% \begin{itemize}
% 	\item RI5CY
% 	\item picorv?
% 	\item Vanilla RISC-V (e.g. RI5CY without hwloops/dsp)?
% 	\item WD Superscalar Core
% \end{itemize}

% \begin{figure}
% 	\centering
% 	\includegraphics[width=\linewidth]{example-image-golden}
% 	\caption{Comparison of kernel execution time when running on a basic RISC-V core, the SweRV \cite{wd2018swerv} super-scalar core, RI5CY, and RI5CY with the \gls{ssr} extension. See \secref{sec:results_perf_core}.}
% 	\label{fig:results_perf_core}
% \end{figure}

% \begin{figure}
% 	\centering
% 	\includegraphics[width=\linewidth]{example-image-golden}
% 	\caption{Comparison of floating-point and integer unit utilization ratios achieved by kernels running on a basic RISC-V core, the SweRV \cite{wd2018swerv} super-scalar core, RI5CY, and RI5CY with the \gls{ssr} extension. See \secref{sec:results_perf_core}.}
% 	\label{fig:results_util_core}
% \end{figure}

% ------------------------------------------------------------------------------
\subsection{Multi-Core Comparison}
\label{sec:results_cluster}

To evaluate the impact in a multi-core setting, we deploy the RI5CY core enhanced with \glspl{ssr} described in \secref{sec:results_core} in a PULP cluster \cite{rossi2017energy}.
The cores operate on a shared \gls{tcdm} which provides single-cycle access latency (see \figref{fig:arch_regfile_dataflow}).
The memory is separated into multiple banks that are individually arbitrated.
Multiple cores accessing the same bank will see one core succeed and the other cores stall for a cycle.

\subsubsection{Performance}
\label{sec:results_cluster_perf}

% \begin{figure}
% 	\centering
% 	\includegraphics[width=\linewidth]{cluster_speedup.pdf}
% 	\caption{Speedup of SSRs versus optimized RI5CY code.}
% 	\label{fig:results_cluster_speedup}
% \end{figure}

% \figref{fig:results_cluster_speedup} shows the speedup achieved by the \gls{ssr} extension in a cluster with four RI5CY cores with four \glspl{fpu}, one per core.
The results presented for a single core in \secref{sec:results_core} translate well into a cluster setting with multiple cores, with the speedup remaining largely the same.
This multi-core environment introduces two new sources of overhead compared to a single-core environment:
\begin{enumerate}
	\item Contention in the memory system.
	While the number of banks in the \gls{tcdm} is chosen to provide approximately twice the bandwidth that the processors can request, accesses into the same memory bank still conflict.
	Depending on the memory access pattern of a kernel, this leads to the cores observing stall cycles on the memory interface.
	In practice we observe that by employing data placement such as non-power-of-two data dimensions and offsets, >80\% of memory accesses are served immediately.
	\item Parallelization and coordination overhead.
	The kernels require additional instructions to subdivide the problem space across the available cores.
	These occur generally outside the hot loop of the computation and are easily amortized over the problem size.
	Some kernels such as the FFT operate in stages, with the cores requiring synchronization after each stage.
	This synchronization is achieved with a dedicated event unit that provides efficient hardware barriers \cite{glaser2019hardware}.
	In practice the overhead of this synchronization is negligible.
\end{enumerate}

\begin{figure}
	\centering
	\includegraphics[width=\linewidth]{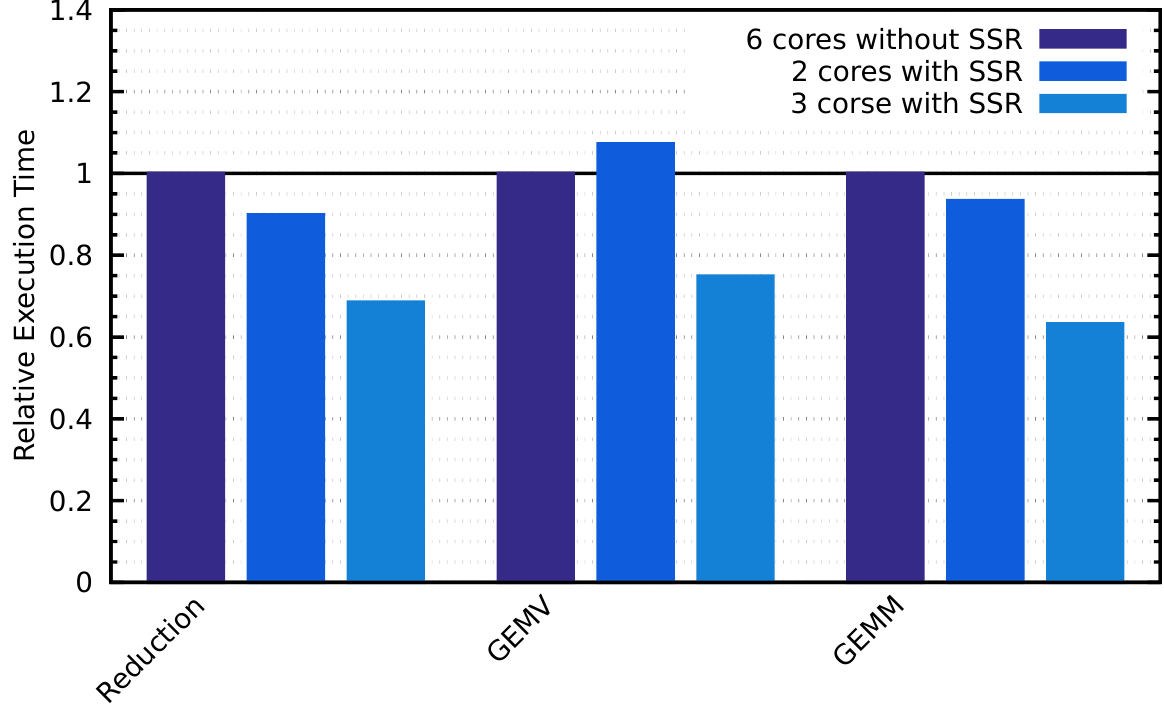}
	\caption{Execution time of a cluster with 3 cores and 3 FPUs with SSRs, a cluster with 2 cores and 2 FPUs with SSRs, both relative to a cluster with 6 cores and 3 FPUs without SSRs.}
	\label{fig:results_cluster_perf}
\end{figure}

Given the two- to three-fold speedup across kernels, we now evaluate by how much the number of cores and FPUs in the \gls{ssr}-augmented cluster can be reduced while maintaining the performance of the non-\gls{ssr} cluster.
To this end we use a six core cluster with three \glspl{fpu}, each shared by two cores, without \glspl{ssr} as performance baseline.
We then evaluate the performance of a two and a three core \gls{ssr}-enabled cluster with core-private \glspl{fpu} to this baseline.
\figref{fig:results_cluster_perf} shows the kernel execution times for the two \gls{ssr} clusters normalized to the execution times in the baseline cluster.
We are interested in the cases where the ratio of execution times is close to one, at which point the reduced cluster perfectly matches the performance of the baseline.
In order to match performance, kernels with an \gls{ssr}-induced speedup around $2\times$ are ideally executed on the three core cluster, while kernels around $3\times$ speedup shall run on the two core cluster.
The next sections evaluate how this reduction in size and complexity translates into area and energy savings.

% This section presents a performance comparison on an entire multi-core cluster, with \gls{tcdm} and ICACHE. Show performance and FPU/ALU utilization gains in such a setup, including the effects of congestions and caching. The following configurations will be compared:

% \begin{itemize}
% 	\item 8 cores, 8 FPUs (private)
% 	\item 8 cores, 4 FPUs (2:1 sharing with APU interconnect, Vega-style)
% 	\item 4 cores, 4 FPUs (private), with SSR
% 	\item 2 cores, 2 FPUs (private), with SSR (should be equivalent to 8C/4FPU config)
% \end{itemize}

\subsubsection{Area}
\label{sec:results_cluster_area}

\begin{figure}
	\centering
	\includegraphics[width=\linewidth]{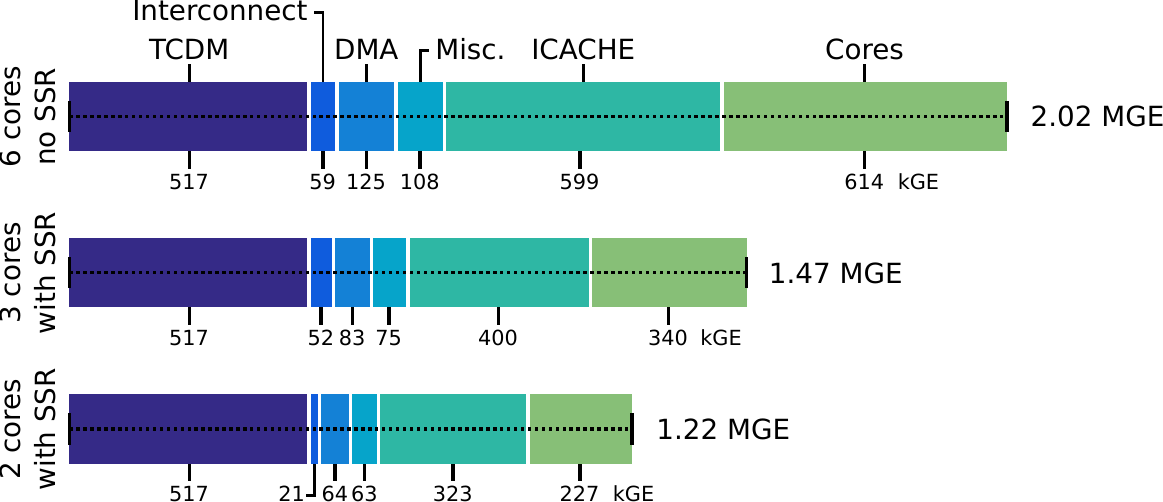}
	\caption{Area breakdown of three clusters: 6 cores/FPUs without SSRs (top), 3 cores/FPUs with SSRs (middle), and 2 cores/FPUs with SSRs (bottom). See \secref{sec:results_cluster_area}.}
	\label{fig:results_cluster_area}
\end{figure}

\figref{fig:results_cluster_area} shows a detailed area breakdown for the three different cluster configurations described in \secref{sec:results_cluster_perf}.
As elaborated above the smaller clusters offer equivalent performance across the kernels due to the addition of \glspl{ssr}.
An interesting observation is the fact that the reduction in the number of cores provides additional savings in the cluster infrastructure.
Fewer cores require less instruction bandwidth, which allows us to reduce the size and parallelism of the instruction cache.
Furthermore the DMA and event unit need to provide fewer control ports.
As such the \SIrange{270}{390}{\kGE} saved by removing cores translate into an overall saving of \SIrange{550}{800}{\kGE} when considering all secondary effects on the cluster.

\begin{figure}
	\centering
	\includegraphics[width=\linewidth]{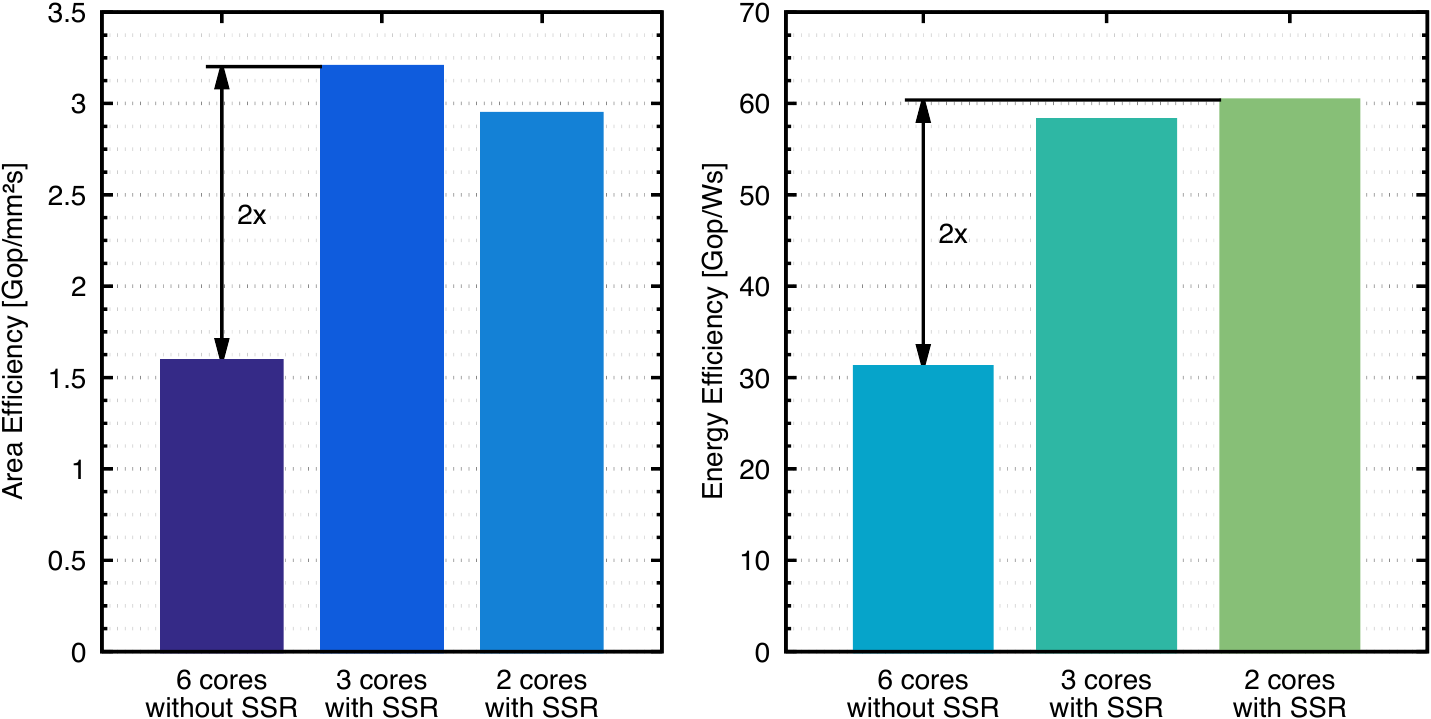}
	\caption{Area efficiency (left) and energy efficiency (right) of the three investigated cluster configurations, given as the number of operations per second achieved on a reduction workload, per area and power consumption of the cluster. See \secref{sec:results_cluster_area} and \secref{sec:results_cluster_power}.}
	\label{fig:results_cluster_eff}
\end{figure}

This reduction in area at no loss in performance translates into a significant efficiency gain in terms of operations per second and area.
\figref{fig:results_cluster_eff} shows the area efficiency achieved for the different cluster configurations.
The introduction of \glspl{ssr} yields and area efficiency improvement of $2\times$.

% - [x] Area savings at constant performance
% - [x] Bang-per-buck in area
% - [x] Unproportional savings due to reduction in ICACHE and interconnect

\subsubsection{Energy Efficiency}
\label{sec:results_cluster_power}

\begin{figure}
	\centering
	\includegraphics[width=\linewidth]{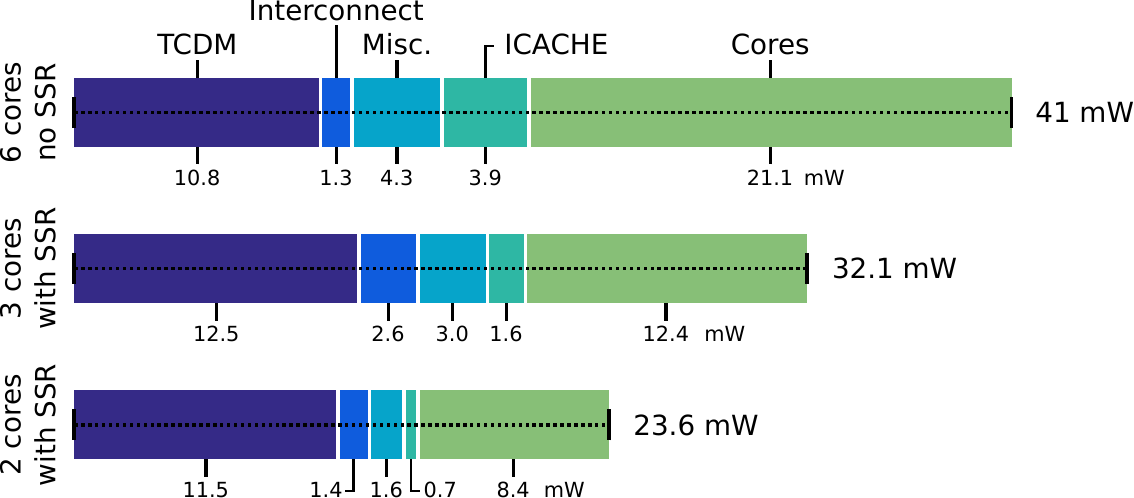}
	\caption{Power breakdown of three clusters: 6 cores/FPUs without SSRs (top), 3 cores/FPUs with SSRs (middle), and 2 cores/FPUs with SSRs (bottom). The clusters are running a reduction kernel; estimated for \SI{1}{\GHz}. See \secref{sec:results_cluster_power}.}
	\label{fig:results_cluster_power}
\end{figure}

\figref{fig:results_cluster_power} shows a power breakdown for the three different cluster configurations.
We use a trace of the reduction kernel to evaluate the power consumption.
Power was estimated for typical silicon at \SI{0.8}{\volt} and \SI{25}{\celsius}.
The smaller clusters with \glspl{ssr} provide a $1.3\times$ and $1.7\times$ reduction in power consumption while maintaining a performance equivalent to the baseline cluster without \glspl{ssr}.
Note again that the reduction in cores saves more than just the core power itself, due to a reduction in cluster infrastructure, as described in \secref{sec:results_cluster_area}.
\figref{fig:results_cluster_eff} shows how the reduction in cluster power influences the per-operation energy efficiency.
The proposed \gls{ssr} extension provides an increase of $2\times$ in energy efficiency across the investigated kernels.
% Across the \SIrange{25}{85}{\celsius} range the proposed \gls{ssr} extension provides an increase of $2\times$ in energy efficiency across the investigated kernels.
% \todo3{In an \gls{hpc} setting at \SI{85}{\celsius} the effect is more pronounced due to a temperature-induced increase in leakage power, with the \glspl{ssr} now achieving a $3\times$ increase in energy efficiency.} Both estimates were performed for typical silicon at \SI{0.8}{\volt}.

\subsubsection{Instruction Pressure}
\label{sec:results_cluster_insts}

One of the main benefits of \glspl{ssr} is that they reduce the number of instruction fetches.
The two-core cluster with \glspl{ssr} executes 3$\times$, 3.2$\times$, and 3.5$\times$ fewer instructions on the kernels in \figref{fig:results_cluster_perf} than the six-core cluster without \glspl{ssr}.
Power consumption of the instruction cache is reduced by 5.6$\times$, further facilitated by the reduced number of cores.

% ------------------------------------------------------------------------------
\subsection{Amdahl's Law and Strong Scaling}
\label{sec:results_amdahl}

We observe an interesting effect of \glspl{ssr} on parallel kernels.
A dot product executed on a single core experiences a speedup of 3$\times$ through the use of \glspl{ssr}.
In a six-core cluster the speedup drops to 2.2$\times$.
This is due to additional overhead such as work splitting and synchronization that is purely sequential and cannot be parallelized, as per Amdahl's law.
As shown in \figref{fig:results_cluster_perf}, a two-core cluster with \glspl{ssr} matches the performance of a six-core cluster without \glspl{ssr}.
This is thanks to the proposed architecture providing a speedup to sequential code by improving how well a single instruction stream can utilize computational resources.
Since this reduces the number of cores --- and thus parallelism --- required to maintain a given compute performance by 2-3$\times$, the implications of Amdahl's law and associated strong scaling are significantly mitigated.
A system with \glspl{ssr} requires 2-3$\times$ less compute resources and separate instruction streams to achieve a given performance, reducing the parallelization overhead correspondingly.

% % ------------------------------------------------------------------------------
% \subsection{Die Power Density}
% \label{sec:results_power_density}

% The improved area and energy efficiency has an interesting effect on power density. In more absolute terms we take a modern V100 GPU as reference \toref3 and assume that in an \gls{hpc} setting we can remove at most \SI{300}{\watt} across a \SI{815}{\milli\meter\square} die, limiting the power density to \SI{0.37}{\watt\per\milli\meter\square} or below.
% At \SI{0.8}{\volt} the dual-core cluster described in \secref{sec:results_cluster} has a power density of \todo3{\SI{0.9}{\watt\per\milli\meter\square}}.
% To lower this value to the above limit we use voltage and frequency scaling, which further improves energy efficiency.
% Previous silicon measurements in 22FDX performed by us showed that in order to achieve the necessary $2.4\times$ power reduction the supply voltage needs to be lowered from \SI{0.8}{\volt} to \todo3{\SI{0.65}{\volt}}.
% This reduction improves energy efficiency by an additional \todo3{$1.5\times$}, which increases the overall energy efficiency gain offered by the proposed \gls{ssr} extension to \todo3{$4.5\times$} across the investigated kernels.

% ------------------------------------------------------------------------------
\subsection{Automated Code Generation}
\label{sec:results_codegen}

% \begin{figure}
% 	\centering
% 	\includegraphics[width=\linewidth]{codegen_quality.pdf}
% 	\caption{Speedup comparison of manually written assembly code leveraging \glspl{ssr} with code generated by LLVM augmented with a \gls{ssr}-specific backend pass.}
% 	\label{fig:results_codegen}
% \end{figure}

% \todo3{Either add more data points and rephrase "programs" to "kernels", or remove the entire section.}
The compiler additions outlined in \secref{sec:progmod_llvm} allow kernels to leverage the \gls{ssr} extension without explicit instruction by the programmer.
Applied to a reduction kernel, our LLVM pass achieves a 2.0$\times$ speedup over the baseline, compared to the 2.1$\times$ speedup of the manual implementation.
In both cases manual replacement of hardware loop and \gls{dsp} instructions were necessary since LLVM does not yet fully support all instruction set extensions of the RI5CY core.
The remaining 5\% of performance gap are due to sub-optimal instruction selection during \gls{ssr} configuration in our prototype pass.
This result shows the promise of using automated passes in later stages of a compiler to leverage \glspl{ssr} in a manner transparent to the programmer.
Existing software would be able to make use of our architectural extension via simple recompilation without the need for re-engineering work.
Further work includes improved handling of nested loops and only selectively mapping loops to \glspl{ssr} based on compile-time heuristics or runtime knowledge.

% \figref{fig:results_codegen} shows that the automatically generated code leveraging the \gls{ssr} extension offers the same performance improvement than manually written assembly across many kernels.
% This makes the proposed architecture transparent to the programmer and allows a wide range of software to automatically benefit from the performance improvement.
% Existing software can make use of our architecture via simple recompilation without the need for re-engineering work.
% Even more so, software that is already written in a vectorization-friendly way to leverage the SIMD capabilities of processors is highly amenable to automated ``streamification'' via \glspl{ssr}.
% Due to the integration into LLVM, use of the extension is not limited to C or C++ but extends to any other language that has an LLVM backend such as C\#, CUDA, Fortran, Haskell, Java, Rust, or Scala, to name a few \toref3.

% ------------------------------------------------------------------------------
\subsection{Comparison to Other Cores}
\label{sec:results_cores}

\begin{table*}
\begin{threeparttable}
\caption{Comparison of the \gls{ssr} architectural extension with other processor cores for a reduction kernel. See \secref{sec:results_cores} for details.}
\label{tbl:results_cores}
\begin{tabularx}{\linewidth}{
	@{}
	X
	l
	l
	l
	S[table-format=2.0]
	S[table-format=3.0]
	S[table-format=4.0]
	S[table-format=2.0]
	S[table-format=1.3]
	S[table-format=3.1]
	S[table-format=2.2]
	S[table-format=2.2]
	@{}
}
\toprule
	\textbf{Core} &
	\textbf{Width} &
	\textbf{Order} &
	\textbf{Issue} &
	{\textbf{Peak Perf.}} &
	{\textbf{Util.}} &
	{\textbf{Freq.}} &
	{\textbf{Tech.}} &
	{\textbf{Area}\tnote{\dag}} &
	{\textbf{Power}} &
	{\textbf{Area Eff.}\tnote{\dag}} &
	{\textbf{Energy Eff.}} \\
 	&
	[bit] &
	&
	&
	{[\si{\op\per\cycle}]} &
	{\textbf{Limit}} &
	{[\si{\MHz}]} &
	{[nm]} &
	{[\si{\milli\meter\squared}]} &
	{[\si{\milli\watt}]} &
	{[\si{\giga\op\per\second\per\milli\meter\squared}]} &
	{[\si{\giga\op\per\second\per\watt}]} \\
\midrule
	2x RI5CY \& SSR [us]                       & 32 & in  & 1x & 2  & 100\,\% &  625 & 22 & 0.243  & 23.6             & 3.20 & 60.4 \\
	6x RI5CY \cite{gautschi2017near}           & 32 & in  & 1x & 6  & 33\,\%  &  625 & 22 & 0.402  & 41.0             & 1.59 & 31.2 \\
	Ariane \cite{zaruba2019cost}               & 64 & in  & 1x & 2  & 33\,\%  & 1700 & 22 & 0.239  & 88.1             & 13.4 & 6.37 \\

	Rocket \cite{lee201445nm}                  & 32 & in  & 1x & 1  & 33\,\%  & 1000 & 40 & 0.118  & 34.0             & 2.80 & 9.71 \\

	BOOM \cite{celio2017boomv2}                & 32 & out & 2x & 2  & 50\,\%  & 1200 & 28 & 0.321  & 118  \,\tnote{*} & 3.74 & 10.1 \\
	SweRV \cite{marena2019riscv} \tnote{\ddag} & 32 & out & 2x & 2  & 50\,\%  & 1000 & 22 & 0.168  & 62.0 \,\tnote{*} & 5.95 & 16.1 \\
	Ara (16 lanes) \cite{cavalcante2019ara}    & 64 & in  & 1x & 32 & 100\,\% & 1040 & 22 & 1.99   & 794              & 15.6 & 41.9 \\
	Hwacha (4 lanes) \cite{lee201445nm}        & 64 & in  & 1x & 8  & 100\,\% & 1000 & 45 & 0.717  & 430              & 11.2 & 18.6 \\

	% ARM Cortex A5 \toref3                               & 32 & in  & 2x & 1  & ?     & ?    & ?  & ?                 \\
	% Intel Core i9-9900K\tnote{*} \cite{schor2019coffee} & 64 & out & 8x & 32 & 25\%  & 3600 & 14 & 20.2\tnote{\ddag} & 4.40\tnote{\ddag} & 1.43 & 6.54 \\
	% Volta V100\tnote{*} \cite{nvidia2017volta}          & -- & in  & 1x & 64 & 33\%  & 1533 & 12 & 14.8\tnote{\ddag} & 1.60\tnote{\ddag} & 2.19 & 20.2 \\
\bottomrule
\end{tabularx}
\begin{tablenotes}
    \item[\dag] Normalized to 22\,nm \cite{brain2017scaling}
	\item[\ddag] Our synthesis in 22\,nm
    \item[*] Area-proportional estimate relative to similar core
    % \item[(?)] \todo3{Should we even show the ``Area Eff.'' column? We don't really have an edge there...}
\end{tablenotes}
\end{threeparttable}
\end{table*}

\tabref{tbl:results_cores} outlines the impact of our architectural extension in comparison to other processor cores.
We count \gls{fma} as \SI{1}{\op} and normalize silicon area to our 22\,nm technology node.
Where otherwise unavailable we have estimated power consumption as proportional to area relative to the Ariane core \cite{zaruba2019cost} listed in \tabref{tbl:results_cores}.
The kernel in question is a reduction with no data reuse, such as a dot product. %, to focus on the instruction issuing bottleneck imposed by explicit load/store instructions.
For our comparison we consider an entire cluster of RI5CY cores as described in \secref{sec:results_cluster} to also capture the contributions of the instruction cache and \gls{tcdm} as an equivalent to the L1 caches in other processors.
In general we observe that due to factors such as bit width, \gls{simd} vectorization, cache sizes, and target speed, area and energy efficiency do not necessarily correlate.

\subsubsection{Peak Utilization}

\tabref{tbl:results_cores} lists the maximum theoretical utilization each core can achieve (``Util. Limit'').
We define this metric as
\begin{equation}
\eta_\text{max} = \lim_{N\to\infty} \eta(N)
\end{equation}
for a problem size $N$.
The case where $N$ tends to infinity is interesting because it captures inherent inefficiencies that remain even when setup and loop overheads have asymptotically disappeared.
For example, a dot product on RI5CY without \glspl{ssr} has a constant setup overhead of two and a loop body of three instructions (see \secref{sec:results_isa_amortization}). With \glspl{ssr} the overhead is seven and the loop body one instruction. Each loop iteration contains one ``useful'' operation. The utilization limit thus is:
\begin{align}
\lim_{N\to\infty} \frac{N}{2+3N} &= 33\% & (\text{without SSR}) \\
\lim_{N\to\infty} \frac{N}{7+N} &= 100\%  & (\text{with SSR})
\end{align}
As shown in \figref{fig:results_amortization}, RI5CY with \glspl{ssr} already reaches close-to-peak utilizations at practical problem sizes, for example 93\% at $N=100$, and 99.3\% at $N=1000$.
This groups processors into ``efficiency classes'' according to the peak utilization they can theoretically achieve.
Single-issue in-order cores are limited to 33\% as elaborated throughout this paper. Dual-issue cores reach 50\% due to the ability to overlap instructions.
Finally vector processors reach up to 100\% due to the ability to overlap long-running instructions.

% Rocket
\subsubsection{Single-Issue Cores}
\label{sec:results_cores_single_issue}
The Rocket core \cite{asanovic2016rocket, lee201445nm} is a single-issue, in-order, 32\,bit core.
It suffers from the same bottleneck as RI5CY \emph{without} \glspl{ssr}.
RI5CY \emph{with} \glspl{ssr} provides a 1.1$\times$ and 6.2$\times$ increase in area and energy efficiency, respectively.

% Boom, SweRV
\subsubsection{Multiple-Issue Cores}
\label{sec:results_cores_multi_issue}
BOOM \cite{celio2017boomv2} and SweRV \cite{marena2019riscv} are both dual-issue, out-of-order, 32\,bit cores.
The ability to issue a load/store in parallel with a computation allows for a peak utilization $\eta$ of 50\% on long reductions.
\glspl{ssr} outperform BOOM by 6$\times$ in terms of energy efficiency, and are roughly on par in terms of area efficiency.
We have synthesized the SweRV core in 22\,nm ourselves, where it reaches \SI{1}{\GHz} in a typical process corner.
Our extension achieves a 1.9$\times$ and 3.8$\times$ higher area and energy efficiency, respectively.
Note that SweRV does not support \gls{fp} instructions in contrast to the other cores, which would further decrease efficiency.

% Ara, Hwacha
\subsubsection{Vector Processors}
\label{sec:results_cores_vector}
In terms of vector processors we consider Ara \cite{cavalcante2019ara} in a 16 lane configuration and Hwacha \cite{lee201445nm} in a four lane configuration.
The design of the RISC-V vector extension allows the vector computation and loads/stores to run in parallel.
This is a similar benefit as with our architecture, allowing the cores to reach a peak utilization $\eta$ of 100\%.
Since both processors operate on 64\,bit data paths, we assume two-way \gls{simd} vectorization of 32\,bit operations by doubling performance and efficiencies.
Our approach provides a 1.4$\times$ and 3.2$\times$ higher energy efficiency.
\glspl{ssr} have a 4.9$\times$ and 3.5$\times$ lower area efficiency, respectively.
The vector processors achieve this with a highly regular data path which entails 32-way and 8-way \gls{simd} vectorization.
This comes at a significant reduction in programmability due to the need for special data placement and shuffling.
\glspl{ssr} on the other hand have no such requirements.

\begin{revised}
More specifically, vector processors are optimized to perform element-wise or reduction operations on long vectors.
This is usually achieved by tightly coupling \gls{fpu} lanes to individual memory banks of the \gls{vrf}, yielding excellent scaling behaviour.
However this also strictly limits which vector elements can form an operand pair for an instruction.
Many interesting problems, e.g., matrix multiplication, FFT, convolution, or stencils, exhibit data reuse.
In the presence of data reuse, a vector element is paired with more than one other elements, in possibly many different \gls{vrf} banks.
Leveraging this reuse in a vector processor requires inter-bank data exchange, and is essential to avoid additional memory accesses, which unnecessarily push a kernel towards the memory-bound region.
Such data exchange is commonly implemented by shuffling instructions and units.
In order to keep up with the \glspl{fpu}, and the likely need to shuffle before every computation in a realistic problem, the shuffle unit must match the bandwidth of the \glspl{fpu}.
Since instructions operate in a register-to-register fashion, this doubles the bandwidth load of the \gls{vrf} and requires the vector processor to sustain simultaneous \gls{fpu} and shuffling operations.

Intuitively, a vector processor \emph{without} shuffling unit is comparable to our system (see \secref{sec:results_cluster}) without the all-to-all memory interconnect.
In this scenario our \gls{ssr}-based cluster has a small power and area overhead of \SIrange{3}{8}{\percent} due to the interconnect.
A more realistic vector processor \emph{with} shuffling unit requires additional \gls{vrf} bandwidth and all-to-all bank connectivity, leading to a power overhead of \SIrange{26}{48}{\percent} compared to the equivalent \gls{ssr}-based system (see \figref{fig:results_cluster_power}).
This does not account for additional load on the integer core when preparing index arrays for the shuffle, or the additional bandwidth required for the index array to be streamed into the shuffling unit.
\Glspl{ssr} essentially provide a \emph{free shuffle} for every data word.
\end{revised}

% ------------------------------------------------------------------------------
\subsection{Comparison to Commercial Processors}
\label{sec:results_commercial}

\begin{table}
\begin{threeparttable}
\caption{Peak efficiency comparison of the \gls{ssr} architectural extension with commercial processor cores. See \secref{sec:results_commercial} for details.}
\label{tbl:results_commercial_cores}
\begin{tabularx}{\linewidth}{
	@{}
	X
	S[table-format=2.0]
	S[table-format=2.0]
	S[table-format=1.2]
	S[table-format=2.1]
	@{}
}
\toprule
	\textbf{Core} &
	{\textbf{Tech.}} &
	{\textbf{SIMD}} &
	{\textbf{Area Eff.}} &
	{\textbf{Energy Eff.}} \\
	& {[nm]} & {} & {[\si{\kilo\op\per\second\per\GE}]} & {[\si{\giga\op\per\second\per\watt}]} \\
\midrule
	3x RI5CY \& SSR [us]                 & 22 & 1  & 1.28 & 58.2 \\
	Cortex A5 \cite{lee201445nm}         & 40 & 1  & 1.24 & 12.5 \\
	Core i9-9900K \cite{schor2019coffee} & 14 & 16 & 1.14 & 26.2 \\
	Volta V100 \cite{nvidia2017volta}    & 12 & 32 & 1.32 & 61.2 \\
\bottomrule
\end{tabularx}
\end{threeparttable}
\end{table}

Another avenue to improve compute unit utilization is explored by superscalar processors \cite{hennessy2011computer} as they are built by Intel or AMD, for example.
Architectural extensions to the simple single-issue core enable multiple instructions to operate in parallel and out of order \cite{asanovic2015berkeley, wd2018swerv}.
This enables address calculation and data movement to occur in parallel to the actual computation, ideally to an extent where the \gls{fpu} utilization reaches 100\%.
Here the von Neumann bottleneck itself is not addressed directly, requiring significant instruction bandwidth and hardware resources to sustain computation.
% Other schemes include sharing of functional units among multiple single-issue processor cores \cite{gautschi201665nm} which leverages the fact that a realistic instruction stream can only utilize around one third an FPU.
We note that all these schemes involve a significant \emph{per \gls{fpu}} increase of hardware resources dedicated to fetching and decoding of instructions, while providing capabilities to compute addresses and perform memory accesses in parallel.
These changes come at a considerable area and energy cost and must be further complemented by the memory system and other parts of the core infrastructure.
\glspl{ssr} offer a more lightweight approach to improve \gls{fpu} utilization which directly addresses the von Neumann bottleneck.
Furthermore \glspl{ssr} may have applicability within superscalar processors as well, acting as alternative or complementary approach to \gls{simd} parallelization.

\tabref{tbl:results_commercial_cores} shows a high level comparison of \glspl{ssr} with commercial processor cores considering only peak metrics.
In contrast to the open source cores in \secref{sec:results_cores}, detailed micro-architectural analyses of commercial cores are not readily available.
We therefore perform our own estimates of area and energy efficiency based on public information on technology, die area, and area ratios evident in die shots.
The area efficiency normalized to gate equivalents is roughly equal among all four cores.
A cluster with three \gls{ssr}-enabled RI5CY cores outperforms an ARM Cortex A5 and an Intel Core i9-9900K by 4.7$\times$ and 2.2$\times$ in terms of energy efficiency, and is roughly on par with an Nvidia Volta V100.
Note that the larger Intel and NVIDIA cores have a significant technology-driven energy efficiency advantage.
They also require substantial SIMD/SIMT vectorization to operate efficiently, which is not needed in our architecture.

\section{Related Work}
\label{sec:relwork}

% ------------------------------------------------------------------------------
\subsection{Streaming Acceleration}
\label{sec:relwork_stream}

\subsubsection{NTX}
\label{sec:relwork_stream_ntx}

The \gls{ssr} extension bears similarity with the approach taken in the ``Network Training Accelerator'' \cite{schuiki2019scalable, schuiki2019ntx}.
It leverages the regularity of nested loops and affine addressing to operate directly on multi-dimensional data in a scratch pad memory.
NTX is designed as a co-processor tightly coupled to but positioned outside of a RISC-V core, requiring explicit operation offloading.
Since it can only execute a single fundamental operation in its innermost loop body, kernels that consist of more than one such operation have to be executed in multiple passes.
Extending NTX to allow for multiple operations would require significant hardware additions:
Each instruction in the loop body would require another configuration cycle by the control processor, or an instruction fetch unit to read instructions from memory directly.
The ability to perform different operations would require an instruction decoding stage, local registers, and an ALU.
In essence such an extension would be akin to developing another instruction set processor.
Rather, our architecture is a generalization of this scheme as we pull the essential address generation and data streaming into a RISC-V core to leverage the already existing infrastructure.
Not only does our approach boost the utilization of a RISC-V core significantly, it does so while \emph{decreasing} the number of fetched instructions at a very small area overhead.
The seamless integration into the register file leaves the \gls{isa} untouched, except for the addition of a single \gls{csr}.

\subsubsection{Out-of-Order Processors}
\label{sec:relwork_stream_ooo}

Wang et al. \cite{wang2019stream} have explored the applicability of stream-based memory accesses to out-of-order processors based on high-level architectural simulations.
They conclude that a significant fraction of loop nests in their investigated benchmarks is amenable to representation as streams, which supports the wide applicability of our architectural extension.
We provide a more hardware-centric view and show that stream semantics can be embedded into an existing \gls{isa} in a very lightweight and non-invasive way.
Furthermore we find that such a streaming extension applies especially well to small, single-issue, in-order cores, where it provides much more significant speedups and energy savings than in out-of-order cores.

\begin{revised}
\subsubsection{WM Machine}
\label{sec:relwork_stream_wm}

Wulf et al. \cite{wulf1988wm} have presented a machine architecture where loads and stores act on \glspl{fifo} rather than directly on a register file.
The \glspl{fifo} are exposed in the \gls{isa} as a special ``register zero'' (\texttt{r0}/\texttt{f0}).
This is similar to our approach.
Yet it requires explicit loads/stores, address calculation, and branching due to the lack of hardware loops, which reduces the positive impact on tight loop performance.
\end{revised}

% ------------------------------------------------------------------------------
\begin{revised}
\subsection{Loop Acceleration}
\label{sec:relwork_loop}

Hayenga et al. \cite{hayenga2014revolver} have targeted loop execution on large-scale, superscalar, out-of-order x86-style processors.
The authors modify such a processor such that loops in the instruction stream pass through the processor frontend only once.
Instructions are then re-issued multiple times in the backend, which is given self-iteration capabilities.
The approach thus reduces frontend overheads which arise in such large-scale processors.
The loop execution and potential to pre-execute loads is conceptually similar to \glspl{ssr}.
Our proposed \gls{ssr} architecture does not focus on large-scale processors, but offers a solution across a potentially wide range of processor sizes.
In contrast to this architecture, \glspl{ssr} support nested loops and allow for loads/stores to be completely elided.
We show that the benefits are significant especially in small-scale in-order processors.
\end{revised}

% ------------------------------------------------------------------------------
\subsection{Data Address Generator}
\label{sec:relwork_dag}

Data address generators have been thoroughly studied in literature \cite{talavera2008address, sudarsanam1999analysis, kuulusa1997flexible}, and are the subject of a wide range of patents \cite{pfeiffer1991high, gove1997address, pekarich2003address}.
Their complexity ranges from simple pointer incrementing functionality to extensions made to accommodate address patterns commonly found in Fourier and discrete cosine transforms \cite{luo1996fast}.

\subsubsection{\glspl{dsp}}
\label{sec:relwork_dag_dsp}

Address generators are well-established in \glspl{dsp} \cite{ilic2011address}.
Specialized load/store instructions on designated address registers allow for pointer arithmetic to be subsumed.
These \glspl{agu} generally support a wide range of patterns, modes, and modulo operations, which makes them costly in terms of area and energy footprint \cite{talavera2008address}.
Our work extends this by four essential insights:
\begin{enumerate}
\item streaming semantics on registers allow for elision of load/store instructions in addition to pointer arithmetic;
\item \glspl{ssr} remove the need for specialized instructions, reducing the \gls{isa} impact to almost zero by allowing all existing instructions to leverage streams and \glspl{agu};
\item this allows for streaming operations and \glspl{agu} to be integrated into small, energy-efficient, general-purpose cores; and
\item a lean and efficient dedicated \gls{agu} for affine address patterns and loop nests is sufficient to provide significant 2x to 3x gains in such cores, while leaving more complex and irregular patterns to be handled by the instruction stream.
As we have shown many kernels can be arranged such that their innermost hot loops are of such a regular structure, and handling of further complexity in the outer loops is amortized well over the computation.
\end{enumerate}

\subsubsection{Superscalar Processors}
\label{sec:relwork_dag_superscalar}

Modern superscalar processors such as those built by Intel and AMD include multiple \glspl{agu} closely coupled to the load/store units to perform address calculation.
These \glspl{agu} may be triggered as part of the indirection in a memory operand as part of a \gls{cisc} instruction, or by using explicit string instructions such as \texttt{REP} with \texttt{LODS/STOS} \cite{van2008address}.
Our proposed \gls{ssr} extension has less impact on the \gls{isa} as it does not require additional, dedicated instructions.
Provisioning for explicit streaming instructions would require a 16x replication of each instruction in a four-operand \gls{isa} such as RISC-V, and 9x in a three-operand \gls{isa} such as x86.
Furthermore we show that it is possible to encode memory accesses in compute instructions even in a \gls{risc} architecture, without the need for complex operand addressing modes.

% ------------------------------------------------------------------------------
\subsection{Vector Processors}

% VectorBlox ORCA LVE
\subsubsection{ORCA LVE}
An approach similar to ours has been proposed in the ``Lightweight Vector Extension'' of the ORCA processor \cite{lemieux2016orca, lemieux2019tinbinn}.
This LVE streams data directly from a scratch pad memory through the ALU of a processor and stores the result back in memory.
One fundamental operation is performed on the data as it is in flight, making the approach similar to NTX \cite{schuiki2019scalable}.
It suffers from the same limitation, requiring multiple ``ping-pong'' passes from and to memory to compute more complex kernels.
In its current implementation, the LVE only supports one-dimensional data.
The \gls{ssr} extension is a superset of the LVE and given the similarity of the approaches should be comparable in hardware complexity.

% Ara and Hwacha
\subsubsection{RISC-V ``V'' Extension}
Other approaches to improve \gls{fpu} utilization exist.
Vector processors have seen an increase in popularity recently with new architectures such as Hwacha \cite{lee201445nm} and Ara \cite{cavalcante2019ara} being proposed, and vector semantics being integrated into \glspl{isa} such as RISC-V.
These processors amortize instruction overheads by encoding identical operations on multiple different operands in a single instruction, allowing a single instruction to trigger hundreds of operations.
Such a scheme requires high operational and spatial regularity as is commonly found in \gls{simd} data paths, but in an amplified fashion due to the increased and often dynamic vector length.
This requires powerful scatter/gather accesses into the memory system and shuffle operations to reorganize data within the register file, which add complexity and do not contribute directly to the computation result.
Our proposed architectural allows instructions to operate directly on the scratch pad memory at word granularity, essentially providing a ``free'' shuffle operation before and after each computation.
The freedom in terms of data access patterns provided by the data mover in our architecture allows for a significantly more relaxed programming model, with less emphasis on data placement and no need for shuffling instructions.
In a sense, an \gls{ssr} machine is a superset of a vector machine in that instructions for the latter can readily be translated into the former, but not vice versa.

% % ==============================================================================
% \section{Future Work}
% \label{sec:futwork}

% \todo3{\begin{itemize}
% 	\item Highly-tuned small core to leverage \glspl{ssr} (hook for Snitch)
% 	\item Scale out study (to compare apples to apples w.r.t. Intel, ARM)
% 	\item Sparsity
% \end{itemize}}

% ==============================================================================
\section{Conclusion}
\label{sec:conc}

In this work we have presented the ``Stream Semantic Register'' \gls{isa} extension for single-issue in-order processor cores.
It introduces stream semantics on a subset of the processor's registers which can be enabled and disabled at the software's discretion via an additional \gls{csr}.
This leaves all existing instructions in the \gls{isa} virtually untouched and allows them to leverage data streams.
\glspl{ssr} bring a 2$\times$ to 5$\times$ speedup at the \gls{isa} level by allowing load/store instructions to be elided.
We have implemented the \gls{ssr} extension into RI5CY, a highly \gls{dsp}-optimized RISC-V core.
In this setting, it provides a 2$\times$ to 3.7$\times$ speedup and 1.5$\times$ to 2.3$\times$ energy efficiency gain across a wide range of data-oblivious kernels.
This comes at a modest cost of a 5\% increase in critical path delay and 11\% increase in area.
In a multi-core cluster the number of cores can be reduced by 2$\times$ to 3$\times$ while maintaining the same performance, which improves area and energy efficiency by 2$\times$.
\glspl{ssr} reduce the number of fetched instructions by 3.5$\times$ and instruction cache power by 5.6$\times$ on realistic kernels.

% - [x] Almost no impact on ISA
% - [x] Wide range of kernels where this is applicable
% - [x] ISA-level speedups
% - [x] Per-core speedups
% - [x] Cluster-level speedups
% - [x] Energy efficiency gain per cluster
% - [x] Reduced instruction fetch
% - [x] Comparison to other cores?

% \todo3{We have presented SSR. It blows our mind. Yours too, hopefully.}

%%%%%%%%%%%%%%%%%%%%%%%%%
%%   Acknowledgments   %%
%%%%%%%%%%%%%%%%%%%%%%%%%

\section*{Acknowledgments}

This work has received funding from the European Union’s Horizon 2020 research and innovation programme under grant agreement number 732631, project ``OPRECOMP''.

% Supported in part by the European Union's H2020 Specific Grant Agreement for European Processor Initiative (EPI) under grant agreement number 826647.

%%%%%%%%%%%%%%%%%%%%%%
%%   Bibliography   %%
%%%%%%%%%%%%%%%%%%%%%%

\bibliographystyle{IEEEtran}
\bibliography{IEEEabrv,bibliography}

%%%%%%%%%%%%%%%%%%%%%
%%   Biographies   %%
%%%%%%%%%%%%%%%%%%%%%

\newpage

\begin{IEEEbiography}[{\includegraphics[width=1in,height=1.25in,clip,keepaspectratio]{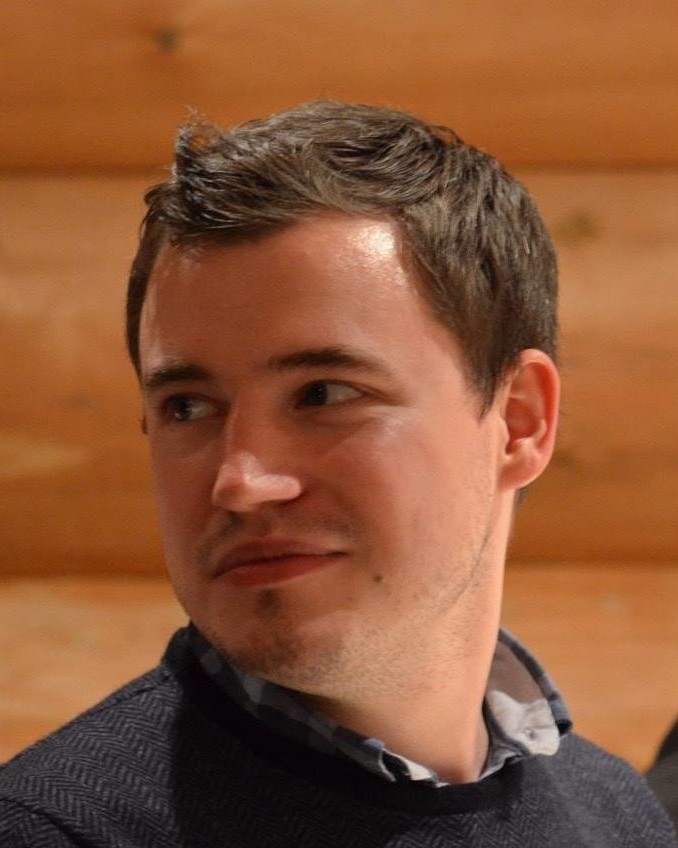}}]{Fabian Schuiki}
  received the B.Sc. and M.Sc. degree in electrical engineering from ETH Zürich, in 2014 and 2016, respectively.  He is currently pursuing a Ph.D. degree with the Digital Circuits and Systems group of Luca Benini.  His research interests include computer architecture, transprecision computing, as well as near- and in-memory processing.
\end{IEEEbiography}

\begin{IEEEbiography}[{\includegraphics[width=1in,height=1.25in,clip,keepaspectratio]{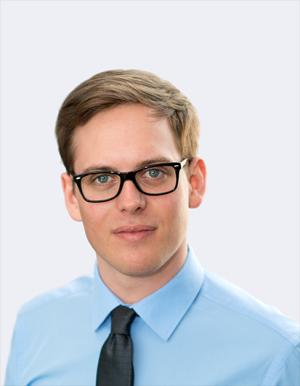}}]{Florian Zaruba}
  received his BSc degree from TU Wien in 2014 and his MSc from the Swiss Federal Institute of Technology Zurich in 2017.  He is currently pursuing a PhD degree at the Integrated Systems Laboratory.  His research interests include design of very large scale integration circuits and high performance computer architectures.
\end{IEEEbiography}

\begin{IEEEbiography}[{\includegraphics[width=1in,height=1.25in,clip,keepaspectratio]{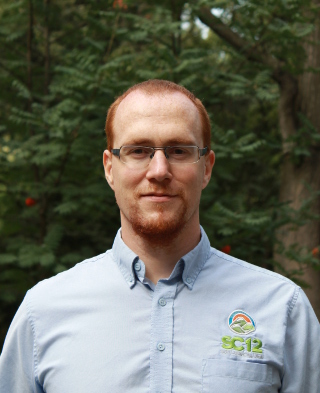}}]{Torsten Hoefler}
  is a Professor of Computer Science at ETH Zürich, Switzerland. He is also a key member of the Message Passing Interface (MPI) Forum where he chairs the ``Collective Operations and Topologies'' working group.  His research interests revolve around the central topic of ``Performance-centric System Design'' and include scalable networks, parallel programming techniques, and performance modeling.  Torsten won best paper awards at the ACM/IEEE Supercomputing Conference SC10, SC13, SC14, EuroMPI'13, HPDC'15, HPDC'16, IPDPS'15, and other conferences.  He published numerous peer-reviewed scientific conference and journal articles and authored chapters of the MPI-2.2 and MPI-3.0 standards.  He received the Latsis prize of ETH Zurich as well as an ERC starting grant in 2015.
  % Additional information about Torsten can be found on his homepage at htor.inf.ethz.ch.
\end{IEEEbiography}

\begin{IEEEbiography}[{\includegraphics[width=1in,height=1.25in,clip,keepaspectratio]{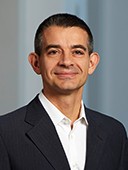}}]{Luca Benini}
  holds the chair of digital Circuits and systems at ETHZ and is Full Professor at the Universita di Bologna.  Dr. Benini's research interests are in energy-efficient computing systems design, from embedded to high-performance.  He has published more than 1000 peer-reviewed papers and five books.  He is a Fellow of the ACM and a member of the Academia Europaea. He is the recipient of the 2016 IEEE CAS Mac Van Valkenburg award.
\end{IEEEbiography}

\vfill

\end{document}